\documentclass[aps,pra,nofootinbib,twocolumn,superscriptaddress]{revtex4-2}
\usepackage[T1]{fontenc}

\usepackage[dvipsnames]{xcolor}
\usepackage{hyperref}
\hypersetup{
  breaklinks=true,
  colorlinks=true,
  allcolors=BlueViolet,
}

\usepackage{physics,braket,bm,amssymb} 
\renewcommand{\t}{\text} 
\newcommand{\f}[2]{\dfrac{#1}{#2}} 
\newcommand{\p}[1]{\left(#1\right)} 
\renewcommand{\sp}[1]{\left[#1\right]} 
\renewcommand{\set}[1]{\left\{#1\right\}} 
\newcommand{\bk}{\braket} 
\renewcommand{\v}{\bm} 
\renewcommand{\dd}{\text{d}} 
\renewcommand{\i}{\mathrm{i}\mkern1mu} 

\newcommand{\bbk}[1]{\langle\!\langle #1 \rangle\!\rangle}

\usepackage{dsfont} 
\newcommand{\1}{\mathds{1}}

\renewcommand{\b}{\hat b}
\renewcommand{\c}{\hat c}
\newcommand{\s}{\hat s}
\renewcommand{\H}{\hat H}
\renewcommand{\S}{\hat S}
\newcommand{\R}{\hat R}
\newcommand{\T}{\hat T}
\renewcommand{\O}{\hat{\mathcal{O}}}
\renewcommand{\P}{\hat{\mathcal{P}}}

\newcommand{\up}{\uparrow}
\newcommand{\dn}{\downarrow}
\newcommand{\x}{\text{x}}
\newcommand{\y}{\text{y}}
\newcommand{\z}{\text{z}}
\newcommand{\g}{\text{g}}
\newcommand{\e}{\text{e}}

\newcommand{\ZZ}{\mathbb{Z}}
\newcommand{\A}{\mathcal{A}}
\newcommand{\C}{\mathcal{C}}
\newcommand{\D}{\mathcal{D}}
\newcommand{\E}{\mathcal{E}}
\newcommand{\I}{\mathcal{I}}

\newcommand{\xx}{\x\x}
\newcommand{\xxi}{\x\x_\i}
\newcommand{\X}{\text{X}}
\newcommand{\XX}{\X\X}
\newcommand{\XXI}{\X\X_\i}
\newcommand{\spin}{\text{spin}}
\newcommand{\eff}{\text{eff}}
\newcommand{\crit}{\text{crit}}
\newcommand{\MF}{\text{MF}}

\newcommand{\sds}{\bar{\v s}\cdot\bar{\v s}}
\DeclareMathOperator{\diag}{diag}

\newcommand{\col}{\underline}
\newcommand{\mean}{\overline}

\usepackage[inline]{enumitem}
\setlist[enumerate,1]{label={(\roman*)}} 

\usepackage{graphicx} 

\newcommand{\JILA}{JILA, National Institute of Standards and Technology and University of Colorado, 440 UCB, Boulder, Colorado 80309, USA}
\newcommand{\CTQM}{Center for Theory of Quantum Matter, University of Colorado, Boulder, CO, 80309, USA}
\newcommand{\UO}{Homer L. Dodge Department of Physics and Astronomy, The University of Oklahoma, Norman, Oklahoma 73019, USA}
\newcommand{\CQRT}{Center for Quantum Research and Technology, The University of Oklahoma, Norman, Oklahoma 73019, USA}

\begin{document}
\interfootnotelinepenalty=10000 

\title{Engineering infinite-range SU($n$) interactions with spin-orbit-coupled fermions in an optical lattice}
\author{Michael A. Perlin}
\email{mika.perlin@gmail.com}
\author{Diego Barberena}
\author{Mikhail Mamaev}
\author{Bhuvanesh Sundar}
\affiliation{\JILA}
\affiliation{\CTQM}
\author{Robert J.~Lewis-Swan}
\affiliation{\UO}
\affiliation{\CQRT}
\author{Ana Maria Rey}
\affiliation{\JILA}
\affiliation{\CTQM}

\date{22 September 2021}

\begin{abstract}
We study multilevel fermions in an optical lattice described by the Hubbard model with on site SU($n$)-symmetric interactions.
We show that in an appropriate parameter regime this system can be mapped onto a spin model with all-to-all SU($n$)-symmetric couplings.
Raman pulses that address internal spin states modify the atomic dispersion relation and induce spin-orbit coupling, which can act as a synthetic inhomogeneous magnetic field that competes with the SU($n$) exchange interactions.
We investigate the mean-field dynamical phase diagram of the resulting model as a function of $n$ and different initial configurations that are accessible with Raman pulses.
Consistent with previous studies for $n=2$, we find that for some initial states the spin model exhibits two distinct dynamical phases that obey simple scaling relations with $n$.
Moreover, for $n>2$ we find that dynamical behavior can be highly sensitive to initial intra-spin coherences.
Our predictions are readily testable in current experiments with ultracold alkaline-earth(-like) atoms.
\end{abstract}

\maketitle

\section{Introduction}
\label{sec:intro}

SU($n$) symmetries play an important role in physics.
Underpinning much of high energy physics, the SU($n$) gauge theory known as Yang-Mills theory is central to our understanding of the electroweak and strong forces.
Extensions of Yang-Mills and SU($n$) symmetry feature in the most well-studied examples of holographic duality \cite{maldacena1999largen} and the connection between entanglement and gravity \cite{ryu2006holographic} through the anti-de Sitter/conformal field theory (AdS/CFT) correspondence.
In a condensed matter setting, SU(2) appears ubiquitously as a symmetry of the Hubbard model, with important consequences for the study of quantum magnetism and high temperature superconductivity \cite{lee2006doping}.
The extension of SU(2) Hubbard and spin models to SU($n$) has led to predictions of exotic phases of matter such as valence bond solids \cite{read1989valencebond, rokhsar1990quadratic, kaul2012lattice, hermele2011topological} and chiral spin liquids \cite{hermele2009mott, hermele2011topological, chen2016syntheticgaugefield, nataf2016chiral}, as well as the potential to perform universal topological quantum computation \cite{freedman2004class, nayak2008nonabelian} and other phenomena \cite{nataf2014exact, nataf2016exact}.
Furthermore, disordered SU($n$) spin models have opened analytically tractable avenues for studying quantum chaos and information scrambling \cite{sachdev1993gapless}.

The tremendous theoretical significance of SU($n$) symmetries makes it all the more exciting that they appear naturally in experimental atomic, molecular, and optical (AMO) platforms with exquisite degrees of microscopic control.
This symmetry arises through the independence of atomic orbital and interaction parameters on the $n$ nuclear spin states of alkaline-earth(-like) atoms, with e.g.~$n=10$ for ${}^{87}$Sr \cite{wu2003exact, cazalilla2009ultracold, gorshkov2010twoorbital, cazalilla2014ultracold}.
As a result, AMO experiments can directly probe the role of SU($n$) interactions in controllable settings.
Recent progress includes studies of the thermodynamic properties of SU($n$) fermionic gases \cite{hazzard2012hightemperature, bonnes2012adiabatic, stellmer2013degenerate, yip2014theory, pagano2014onedimensional, choudhury2020collective, song2020evidence, sonderhouse2020thermodynamics}, SU($n$) Hubbard phases and phase transitions \cite{taie2012su, hofrichter2016direct, taie2020observation}, single- \cite{messio2012entropy} and two-orbital \cite{cappellini2014direct, scazza2014observation, zhang2014spectroscopic, beverland2016realizing} SU($n$) magnetism, and multi-body SU($n$)-symmetric interactions \cite{goban2018emergence, perlin2019effective}.

In the spirit of quantum simulation, further investigations in controlled settings will play an important role in understanding the consequence of SU($n$) symmetries for fundamental questions in physics, as well as their practical use in technological applications.
For example, SU(2)-symmetric spin interactions can be harnessed to develop quantum sensors that surpass classical limits on measurement precision \cite{he2019engineering, perlin2020spin}.
The prospect of similarly exploiting more general SU($n$) symmetries to achieve a technological advantage is still an unexplored avenue of research with untapped potential.

In this work, we consider an experimentally relevant and theoretically tractable regime of the SU($n$) Hubbard model, highlighting differences and similarities with the more familiar case of SU(2).
Working at ultracold temperatures and unit spatial filling (one atom per lattice site), we begin by mapping the SU($n$) Hubbard model onto a multilevel spin model with all-to-all SU($n$)-symmetric interactions in Section \ref{sec:spin_model}.
In Section \ref{sec:controls} we consider the use of control fields to address nuclear spins, finding a simple three-laser driving scheme that allows for the preparation of interesting states with nontrivial intra-spin correlations when $n>2$.
We consider the effect of spin-orbit coupling (SOC) induced by control fields in Section \ref{sec:SOC}, finding in particular that the weak-SOC limit generally gives rise to a (synthetic) inhomogeneous magnetic field, extending previously known results to $n>2$ \cite{mancini2015observation, wall2016synthetic, livi2016synthetic, kolkowitz2016spinorbitcoupled, bromley2018dynamics, he2019engineering}.
Finally, we combine these ingredients to examine mean-field dynamical behaviors of the SU($n$) spin model in Section \ref{sec:mean_field}, finding that:
\begin{enumerate*}
\item long-time-averaged observables obey simple scaling relations with the spin dimension $n$, exhibiting (for spin-polarized initial states) dynamical ferromagnetic and dynamical paramagnetic phases, as previously seen for the case of $n=2$ \cite{smale2019observation, lewis-swan2021cavityqed}, and
\item for $n>2$ the long-time dynamics can be highly sensitive to the intra-spin coherences of the initial state.
\end{enumerate*}
We conclude and discuss future directions in Section \ref{sec:conclusions}.

\section{From lattice fermions to an SU($n$) spin model}
\label{sec:spin_model}

Here we derive a collective SU($n$) spin model for a system of ultracold alkaline-earth(-like) atoms trapped in an optical lattice.
Without external driving fields, the evolution of such atoms in their electronic ground state is governed by the single-body kinetic and two-body interaction Hamiltonians
\begin{align}
  \H_{\t{kin}}
  &= -J \sum_{\bk{j,j'},\mu} \c_{j\mu}^\dag \c_{j'\mu} + \t{h.c.},
  \label{eq:H_kin_FH} \\
  \H_{\t{int}}
  &= \f{U}{2} \sum_{j,\mu,\nu}
  \c_{j\mu}^\dag \c_{j\mu} \c_{j\nu}^\dag \c_{j\nu},
  \label{eq:H_int_FH}
\end{align}
where $\bk{j,j'}$ denotes neighboring lattice sites $j$ and $j'$; $\mu,\nu\in\set{s,s-1,\cdots,-s}$ index orthogonal spin states of a spin-$s$ nucleus, with $s=\frac{n-1}{2}$ (e.g.~$s=\frac{9}{2}$ in the case of ${}^{87}$Sr with $10$ nuclear spin states); $\c_{j\mu}$ is a fermionic annihilation operator, $J$ is a tunneling amplitude (for simplicity assumed to be the same in all directions); and $U$ is a two-body on-site interaction energy.
In the present work, we neglect inter-site interactions and interaction-assisted hopping, which is a good approximation for a sufficiently deep lattice, namely when $J\lesssim E_{\t{R}}$, where $E_{\t{R}}$ is the atom recoil energy.
For simplicity, we now assume a one-dimensional periodic lattice of $L$ sites, and expand the on-site fermionic operators in terms of operators addressing (quasi-)momentum modes $q$ (in units with lattice spacing $a=1$), $\c_{j\mu} = \frac1{\sqrt{L}} \sum_q e^{-\i q\cdot j} \c_{q\mu}$, finding that
\begin{align}
  \H_{\t{kin}}
  &= -2J\sum_{q,\mu} \cos\p{q} \c_{q\mu}^\dag \c_{q\mu}, \\
  \H_{\t{int}}
  &= \f{u}{2N} \sum_{k,\ell,p,q,\mu,\nu}
  \c_{k\mu}^\dag \c_{\ell\mu} \c_{p\nu}^\dag \c_{q\nu}
  \times \delta_{k+p,\ell+q},
  \label{eq:H_int_momenta}
\end{align}
where $N$ is the total number of atoms on the lattice, we define $u\equiv U\times N/L$ for convenience, $\delta_{k+p,\ell+q}=1$ if $k+p=\ell+q$ and zero otherwise (enforcing conservation of momentum).

If the interaction energy $U$ is small compared to the single-particle bandwidth $4J$, then the mode-changing collisions in $\H_{\t{int}}$ become off-resonant, motivating the frozen-mode approximation $\set{k,p}=\set{\ell,q}$ (i.e.~either $k=\ell$ and $p=q$, or $k=q$ and $p=\ell$)\footnote{Note that the frozen-mode approximation neglects correlated momentum-hopping terms of the form $\c_{\pi-p,\mu}^\dag \c_{\pi-q,\mu} \c_{p\nu}^\dag \c_{q\nu}$, which conserve both momentum and energy.
We defer a careful treatment of these terms to future work, noting only that they vanish on the manifold of permutationally symmetric spin states with one atom per lattice site, and that the frozen-mode approximation is benchmarked in Refs.~\cite{he2019engineering, smale2019observation} and Appendix \ref{sec:benchmarking}.}.
The terms with $k=\ell$ and $p=q$ are $\frac{u}{2N} \sum \c_{k\mu}^\dag \c_{k\mu} \c_{p\nu}^\dag \c_{p\nu} = \frac12 N u$, which is a constant  energy shift that we can freely neglect.
Defining the spin operators $\s_{\mu\nu q}\equiv \c_{q\mu}^\dag \c_{q\nu}$, the remaining terms of the kinetic and interaction Hamiltonians are
\begin{align}
  \H_{\t{kin}}
  &= -2J\sum_{q,\mu} \cos\p{q} \s_{\mu\mu q},
  \label{H_kin_start} \\
  \H_{\t{int}}
  &= -\f{u}{2N} \sum_{p,q,\mu,\nu} \s_{\mu\nu p} \s_{\nu\mu q}.
  \label{eq:H_int_start}
\end{align}
Throughout this work, we will assume that atomic modes are singly-occupied, e.g.~due to the initialization of a spin-polarized state with one atom per lattice site, in which multiple occupation of an atomic mode is forbidden by fermionic statistics (Pauli exclusion).
In this case we can simply treat our system as $N$ distinguishable $n$-level quantum spins at ``lattice sites'' $p,q$.
Note that the ``kinetic'' terms of this spin model ($\H_{\t{kin}}$) are proportional to the identity operator, contributing an overall shift in energy that we can neglect at this point.
Nevertheless, these kinetic terms will become important in the presence of an external drive, which we discuss in Section \ref{sec:SOC}.
The validity of approximating the Hubbard model in Eqs.~\eqref{eq:H_kin_FH}--\eqref{eq:H_int_FH} by the spin model in Eqs.~\eqref{H_kin_start}--\eqref{eq:H_int_start} has been previously benchmarked for SU(2)-symmetric interactions \cite{he2019engineering, smale2019observation}, and we provide additional benchmarking for SU(4) and SU(6) in Appendix \ref{sec:benchmarking}.

To further simplify the interaction Hamiltonian $\H_{\t{int}}$ and write it in a form reminiscent of more familiar SU(2) spin models, we now construct the operator-valued spin matrix
\begin{align}
  \v\s_q \equiv \sum_{\mu,\nu} \s_{\mu\nu q} \op{\mu}{\nu},
\end{align}
and for any pair of such operator-valued matrices $\v{\hat{A}},\v{\hat{B}}$, we define the inner product
\begin{align}
  \v{\hat{A}} \cdot \v{\hat{B}}
  \equiv \sum_{\mu,\nu} \hat A_{\mu\nu}^\dag \hat B_{\mu\nu}.
  \label{eq:dot_product}
\end{align}
These definitions allow us to write the spin Hamiltonian in Eq.~\eqref{eq:H_int_start} as
\begin{align}
  \H_{\t{int}} = -\f{u}{2N} \sum_{p,q} \v\s_p\cdot\v\s_q
  = -\f{u}{2N}\v\S\cdot\v\S,
  \label{eq:H_int}
\end{align}
where $\v\S\equiv\sum_q\v\s_q$ is a collective spin matrix, analogous to the collective spin vector $\vec S=(\S_\x,\S_\y,\S_\z)$ in the case of SU(2) \cite{he2019engineering}, with $\frac12\v\S\cdot\v\S \simeq \vec S\cdot\vec S = \S_\x^2 + \S_\y^2 + \S_\z^2$ when $n=2$ (here $\simeq$ denotes equality up to identity terms).

\begin{figure}
\centering
\includegraphics[width=\linewidth]{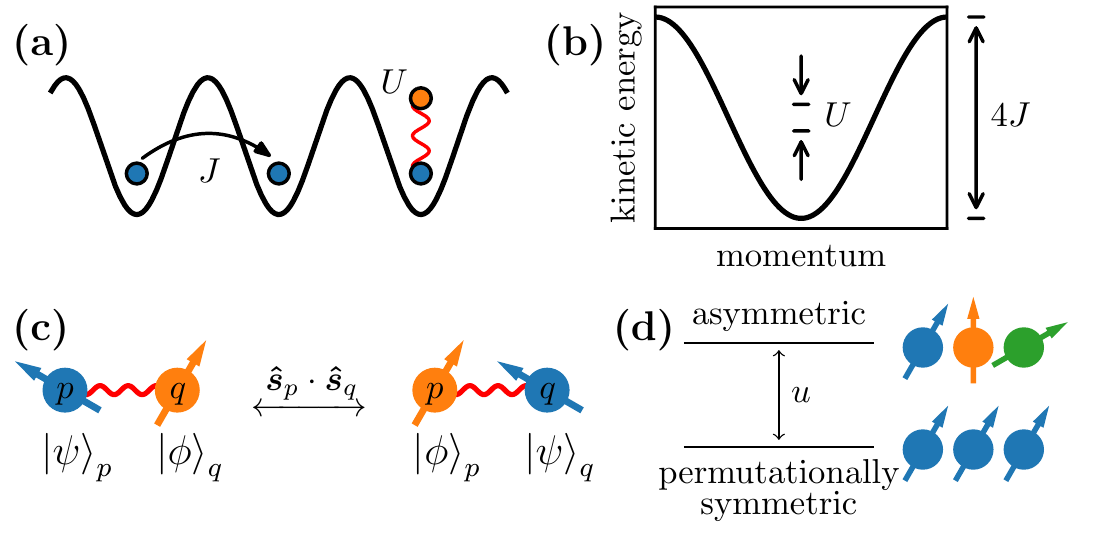}
\caption{
(a) Ultracold atoms on a lattice of $L$ sites tunnel between neighboring lattice sites at a rate $J$, and locally repel each other with interaction energy $U$.
(b,c) When the interaction energy $U$ is small compared to the single-particle bandwidth $4J$, the frozen-mode approximation enables the interaction Hamiltonian to be written as a spin model consisting of exchange terms $\v\s_p\cdot\v\s_q$, which swap the states of two spins pinned to modes $p,q$.
(d) Interactions open an energy gap $u=U\times N/L$ between the manifold of permutationally symmetric states of $N$ spins, and the orthogonal complement of states that break spin-permutation symmetry.
}
\label{fig:spin_model}
\end{figure}

We now discuss the spin Hamiltonian $\H_{\t{int}}$ in Eq.~\eqref{eq:H_int}.
The operator $\v\s_p\cdot\v\s_q$ simply swaps the nuclear spin states of two atoms pinned to modes $p,q$.
The term $-\v\s_p\cdot\v\s_q$ thereby assigns a definite energy of $-1$ ($+1$) to a pair of spins that are symmetric (anti-symmetric) under exchange.
In this sense, $\v\s_p\cdot\v\s_q$ is analogous to the enforcement of SU(2) spin alignment by ferromagnetic interactions, which similarly assigns different energies to the anti-symmetric spin-0 singlet $\ket{\up\dn}-\ket{\dn\up}$ and the symmetric spin-1 triplets $\set{\ket{\up\up},\ket{\dn\dn},\ket{\up\dn}+\ket{\dn\up}}$.
By summing over all pair-wise exchange terms $\v\s_p\cdot\v\s_q$, the interaction Hamiltonian $\H_{\t{int}}$ energetically enforces a permutational symmetry among all spins, opening an energy gap $u$ between the manifold of all permutationally symmetric (PS) states and the orthogonal complement of excited (e.g.~spin-wave) states that break permutational symmetry.
See Figure \ref{fig:spin_model} for a summary of this section thus far.

In the case of SU(2), the PS manifold is precisely the Dicke manifold of collective states $\ket{m_\z}$ with total spin $S=\frac{N}{2}$ and definite spin projection $m_\z\in\set{S,S-1,\cdots,-S}$ onto a fixed quantization axis.
Equivalently, Dicke states $\ket{m_\z}=\ket{m_\up,m_\dn}$ can be labeled by a definite number of spins $m_\up=S+m_\z$ ($m_\dn=S-m_\z$) pointing up (down) along the spin quantization axis, with $m_\up+m_\dn=N$.
In the general case of SU($n$), the PS manifold is similarly spanned by states $\ket{m_s,m_{s-1},\cdots,m_{-s}}$ with a definite number $m_\mu$ of spins in state $\mu$, and $\sum_\mu m_\mu=N$.
The dimension of the PS manifold is equal to the number of ways of assigning $N$ identical spins to $n$ distinct internal states, or ${N+n-1 \choose n-1} \sim N^{n-1}$.

External fields or additional interactions that respect permutational symmetry can induce nontrivial dynamics within the PS manifold.
Moreover, additional terms that explicitly break permutational symmetry can nevertheless lead to interesting dynamics that can be captured within the PS manifold perturbatively, as long as the coupling to non-PS states is weak compared $u$ (see Appendix \ref{sec:pert_theory}) \cite{bravyi2011schrieffer}.
This perturbative regime is thereby efficiently simulable, as the PS manifold has dimension $\sim N^{n-1}$ (as compared to $n^N$ for the entire spin Hilbert space).
Simulating dynamics within the PS manifold requires calculating matrix elements $\bk{\ell|\O|m}$ of spin operators $\O$ with respect to PS states $\ket\ell,\ket{m}$; we discuss this calculation in Appendix \ref{sec:PS_ops}.

\begin{figure}
\centering
\includegraphics[width=\linewidth]{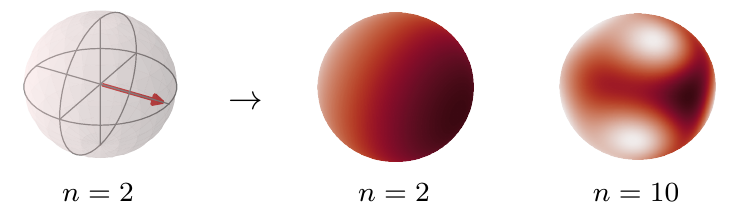}
\caption{
Whereas the state of a two-level spin (qubit) can be represented by a point on (or inside) the Bloch sphere, the state of an $n$-level spin is more generally represented by a probability distribution on the Bloch sphere.
The distribution shown for $n=10$ corresponds to a Haar-random pure state.
}
\label{fig:spin_dist}
\end{figure}

Finally, we take a moment to discuss individual $n$-level spins.
The state of a two-level spin, or a qubit, is commonly represented by a point on (or within) the Bloch sphere.
More generally, the state $\ket\psi$ of an $n$-level spin can be represented by a quasi-probability distribution $Q_\psi$ on the Bloch sphere (commonly known as the Husimi-$Q$ function, e.g.~in the spin-squeezing community \cite{ma2011quantum}).
The value $Q_\psi\p{\v v}$ at a point $\v v$ on the sphere is equal to the overlap of $\ket\psi$ with a pure state $\ket{\v v}$ that is maximally polarized in the direction of $\v v$: $Q_\psi\p{\v v}\equiv\abs{\bk{\v v|\psi}}^2$ (see Figure \ref{fig:spin_dist}).
In the case of a mixed state $\hat\rho$, this distribution is defined by $Q_{\hat\rho}\p{\v v}\equiv\bk{\v v|\hat\rho|\v v}$.
Closely related spherical representations of multilevel spin states and operators are discussed in Refs.~\cite{dowling1994wigner, li2013weylwignermoyal}.
In practice, it is conceptually useful to identify the Hilbert space of a single $n$-level spin with the Dicke manifold of $n-1$ spin-$\frac12$ particles.

\section{External control fields}
\label{sec:controls}

We now consider the addition of external control fields to address atoms' internal spin states, which will determine the observables we can access and initial states we can prepare.
Specifically, we consider off-resonantly addressing an electronic $\ket\g\to\ket\e$ transition of the atoms, and then perturbatively eliminating electronic $\ket\e$ excitations to arrive at an effective ground-state Hamiltonian addressing nuclear spins.
For simplicity, we will assume that the total spin $s$ of the ground- and excited-state (hyperfine) manifolds are the same, as e.g.~with the ${^1}\t{S}_0\to{^3}\t{P}_0$ transition of alkaline-earth-like atoms (AEAs).
However, the results of this section (namely the general form of effective nuclear spin Hamiltonians, as well as the corresponding set of accessible observables and initial states) are the same for transitions that take $s\to s\pm1$, so in practice one is free to address the hyperfine manifolds of the ${^1}\t{S}_0\to{^3}\t{P}_1$ transition of AEAs.

\begin{figure}
\centering
\includegraphics[width=\linewidth]{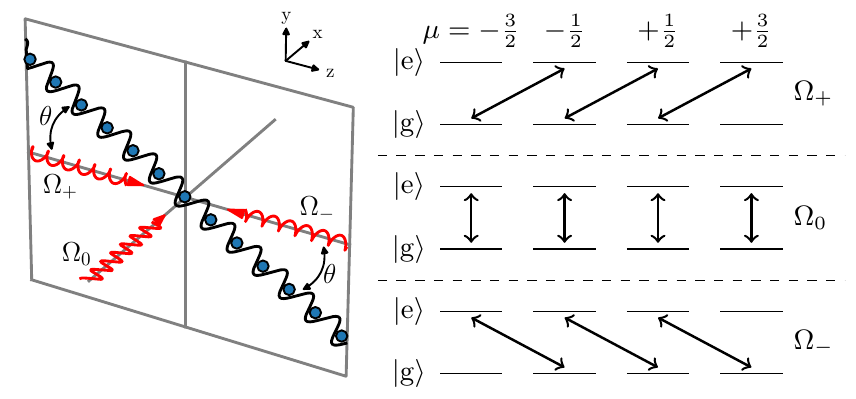}
\caption{
Sketch of the three-laser drive used to address nuclear spins on a one-dimensional lattice.
Two counter-propagating lasers with right-circular polarization and amplitudes $\Omega_\pm$ point at an angle $\theta$ to the lattice axis.
A third, linearly polarized laser with amplitude $\Omega_0$ points in a direction orthogonal to both the lattice and the other driving lasers.
Absorbing a photon from the laser with amplitude $\Omega_m$ induces a transition $(\g,\mu)\to(\e,\mu+m)$ for the (electronic, nuclear spin) state of an atom, where nuclear spin is quantized along the $z$ axis.
}
\label{fig:3LD}
\end{figure}

We consider a specific three-laser driving scheme with a geometry sketched in Figure \ref{fig:3LD}.
Here the lattice lies in the $y$-$z$ plane at an angle $\theta$ to the $z$ axis, oriented along $\v\ell=(0,\sin\theta,\cos\theta)$.
We set the spin quantization axis along $z$.
The laser setup consists of
\begin{enumerate*}
\item two counter-propagating right-circularly polarized lasers with drive amplitudes $\Omega_\pm$ and wavevectors $\kappa\v v_\pm$, propagating in opposite directions along the $z$ axis, $\v v_\pm=(0,0,\pm1)$, and
\item a third laser linearly polarized along $z$, with drive amplitude $\Omega_0$ and wavevector $\kappa \v v_0$, propagating along the $x$ axis $\v v_0=(1,0,0)$.
\end{enumerate*}
All driving lasers are detuned by $\Delta$ below an electronic transition.
The full Hamiltonian for this three-laser drive can be written as
\begin{align}
  \H_{\t{3LD}}^{\t{full}}
  = \sum_{j,m} \Omega_m
  \p{e^{-\i m\phi j} \s_{mj} \otimes\op{\e}{\g}_j + \t{h.c.}}
  + \Delta \hat N_\e,
\end{align}
where $m\in\set{+1,0,-1}$ indexes the laser pointing along $\v v_m$; the SOC angle $\phi\equiv\kappa\v v_+\cdot\v\ell=\kappa \cos\theta$ (in units with lattice spacing $a=1$); $\s_{\z,j},\s_{+,j}\s_{-,j}$ are standard axial, spin-raising, and spin-lowering operators for the spin at lattice site $j$; $\s_{0,j}\equiv\s_{\z,j}$ for shorthand; $\ket\g_j$ and $\ket\e_j$ respectively denote the ground and excited electronic states of atom $j$; and $\hat N_\e=\1\otimes\sum_j\op\e_j$ counts the number of excited atoms (with $\1$ the identity operator on all spin degrees of freedom).

In the far-detuned limit $\abs{\Delta}\gg\abs{\Omega_m}$, a second-order perturbative treatment of electronic excitations ($\ket\e$) yields an effective drive Hamiltonian that only addresses ground-state nuclear spins.
After additionally making the gauge transformation $\s_{mj}\to e^{\i m\phi j}\s_{mj}$ (equivalently $\c_{j\mu}^\dag\to e^{\i\phi\mu j} \c_{j\mu}^\dag$), the drive Hamiltonian then becomes
\begin{align}
  \H_{\t{3LD}} = \sum_j \H_{\t{3LD},j}^{\t{single}},
  \label{eq:H_3LD}
\end{align}
where $\H_{\t{3LD},j}^{\t{single}}$ denotes the action of $\H_{\t{3LD}}^{\t{single}}$ on spin $j$:
\begin{multline}
  \H_{\t{3LD}}^{\t{single}}
  = \tilde\Omega_+ \tilde\Omega_- \s_\z
  + \tilde\Omega_0 \tilde\Omega_- \s_\x
  + \tilde\Omega_0 \tilde\Omega_+ (\s_\z \s_\x  + \s_\x \s_\z) \\
  - \tilde\Omega_0^2 \s_\z^2 - \tilde\Omega_+^2 \s_\x^2
  - \tilde\Omega_-^2 \s_\y^2,
  \label{eq:H_3LD_single}
\end{multline}
with
\begin{align}
  \tilde\Omega_0 \equiv -\f{\Omega_0}{\sqrt\Delta},
  &&
  \tilde\Omega_\pm \equiv \f{\Omega_+\pm\Omega_-}{\sqrt\Delta},
\end{align}
where we have made the simplifying assumption that all drive amplitudes are real to arrive at the form of $\H_{\t{3LD}}^{\t{single}}$ in Eq.~\eqref{eq:H_3LD_single}.
We relax the assumption of real drive amplitudes in Appendix \ref{sec:full_drive}.

\begin{table}
\centering
\caption{
Drive Hamiltonians (left column) that can be implemented with different amplitude-matching conditions (right three columns), some of which are specified by an arbitrary sign $\sigma\in\set{+1,-1}$.
The drives shown here are equal to that of Eq.~\eqref{eq:H_3LD_single} up to a possible energy shift of $\s_\x^2+\s_\y^2+\s_\z^2=s(s+1)$, and come in mutually commuting pairs: a drive with $\abs*{\Omega_m}=1$ and $\Omega_n=0$ for both $n\ne m$ commutes with the drive in which $\Omega_m=0$ and both $\abs*{\Omega_n}=1$.
}
\vspace{.5em}
\begin{tabular}{c||c|c|c}
  $\H_{\t{drive}}^{\t{single}}$
  & $\tilde\Omega_0$ & $\tilde\Omega_+$ & $\tilde\Omega_-$
  \\ \hline\hline
  $-\s_\z^2$ & 1 & 0 & 0
  \\ \hline
   $-\s_\x^2$ & 0 & 1 & 0
  \\ \hline
  $-\s_\y^2$ & 0 & 0 & 1
  \\ \hline
  $\sigma \s_\z + \s_\z^2$ & 0 & 1 & $\sigma$
  \\ \hline
  $\sigma \s_\x + \s_\x^2$ & 1 & 0 & $\sigma$
  \\ \hline
  $\sigma\p{\s_\z \s_\x+\s_\x \s_\z} + \s_\y^2$ & 1 & $\sigma$ & 0
  \\ \hline
  $\pm \s_\z \pm \sigma \s_\x + \sigma (\s_\z \s_\x + \s_\x \s_\z)$
  & 1 & $\sigma$ & $\pm\sigma$
\end{tabular}
\label{tab:drives}
\end{table}

There are three important observations to make about Eqs.~\eqref{eq:H_3LD} and \eqref{eq:H_3LD_single}.
First, the fact that $\H_{\t{3LD}}$ acts identically on all spins means we can freely replace the site index $j$ with a momentum index $q$ (as can be verified by substituting $\c_{j\mu}=\frac1{\sqrt{L}}\sum_k e^{-\i q\cdot j} \c_{q\mu}$), which is important to ensure that this drive addresses the same spin degrees of freedom as the spin Hamiltonians previously considered in Section \ref{sec:spin_model}.
Second, each of $\tilde\Omega_0,\tilde\Omega_+,\tilde\Omega_-$ can be tuned independently by changing the amplitudes of the driving lasers; some particular Hamiltonians for specific values of these amplitudes are shown in Table \ref{tab:drives}.
Third, due to the appearance of mutually commuting pairs of Hamiltonians in Table \ref{tab:drives}, specifically $-\s_\alpha^2$ and $\pm \s_\alpha+\s_\alpha^2$ for $\alpha\in\set{\z,\x}$, the three-laser drive admits pulse sequences that exactly implement arbitrary SU(2) (spatial) rotations of the form $e^{-\i\chi\vec n\cdot\vec s}$, where $\chi$ is a rotation angle, $\vec n$ is a rotation axis, and $\vec s\equiv(\s_\x,\s_\y,\s_\z)$.
The capability to perform arbitrary spatial rotations, together with the capability to measure the number of atoms with spin projection $\mu$ onto a fixed quantization axis, $\bk{\S_{\mu\mu}}$ (where $\S_{\mu\nu}=\sum_j\s_{\mu\nu j}$), implies the capability to reconstruct all components of the mean collective spin matrix $\bk{\v\S}=\sum_{\mu\nu}\bk{\S_{\mu\nu}}\op{\mu}{\nu}$ via spin qudit tomography \cite{newton1968measurability, perlin2020qudit}.
Moreover, we expect that advanced quantum control techniques (similar to those of Refs.~\cite{anderson2015accurate, lucarelli2018quantum}) can be used to implement arbitrary SU($n$) rotations by designing suitable time-dependent drive amplitudes.

If the excited-state manifold $\ket\e$ has total spin $s\pm1$, the effective ground-state Hamiltonians in Eq.~\eqref{eq:H_3LD_single} and Table \ref{tab:drives} remain almost identical, but with some additional $n$-dependent factors that do not affect the general results and discussions above.
These results still hold if (for example) all excited hyperfine manifolds of an electronic ${^1}\t{S}_0\to{^3}\t{P}_1$ transition (with total spins $s+1,s,s-1$) are addressed simultaneously.
See Appendix \ref{sec:full_drive} for additional details.

Finally, we comment on the preparation of initial states.
Initial states are nominally prepared in the ``lab frame'', and must be transformed according to the gauge transformation $\c_{j\mu}^\dag\to e^{\i\phi\mu j} \c_{j\mu}^\dag$ prior to evolution under the three-laser drive $\H_{\t{3LD}}$ in Eq.~\eqref{eq:H_3LD}, which is expressed in the ``gauge frame''.
We assume the capability to prepare an initial state in which all spins are maximally polarized along the $z$ axis, i.e.~$\ket{\z}^{\otimes N}=\ket{s}^{\otimes N}$, which is unaffected by the gauge transformation (up to a global phase).
The three-laser then allows us to rotate this state into one that is polarized along any spatial axis (in the gauge frame).
In addition, when $n>2$ the three-laser drive allows us to prepare product states with nontrivial intra-spin correlations.
For example, when $n$ is even we can prepare an $N$-fold product of the ``kitten'' state
\begin{align}
  e^{-\i\frac{\pi}{2}\p{\s_\y + \s_\y^2}} \ket{s}
  \stackrel{n\,\t{even}}{\propto} \ket{s} + \ket{-s}.
\end{align}
This state has a vanishing mean spin vector, $\bk{\s_\x}=\bk{\s_\y}=\bk{\s_\z}=0$, but variances $\bk{\s_\x^2}=\bk{\s_\y^2}=s/2$ and $\bk{\s_\z^2}=s^2$.

\section{Spin-orbit coupling}
\label{sec:SOC}

We now consider the effect of spin-orbit coupling (SOC) induced by the control fields in Section \ref{sec:controls}.
Before discussing SOC for $n$-level fermions, we briefly review the well-studied case of two-level SOC with a one-dimensional lattice \cite{wall2016synthetic, kolkowitz2016spinorbitcoupled, bromley2018dynamics, he2019engineering}.
In this case, SOC is induced by an external driving field that imprints a phase $e^{-\i\phi j}$ on lattice site $j$, or equivalently imparts a momentum kick $q\to q+\phi$, upon the absorption of a photon\footnote{In order for the drive Hamiltonian $\H_{\t{drive}}^{(\phi)}$ to be well-defined, $\phi$ should be commensurate with the lattice, e.g.~$\phi\in\ZZ\times2\pi/L$ on a one-dimensional lattice of $L$ sites.}:
\begin{align}
  \H_{\t{drive}}^{(\phi)}
  = \f{\Omega}{2} \sum_q \c_{q+\phi,\up}^\dag \c_{q,\dn} + \t{h.c.}.
  \label{eq:drive_2}
\end{align}
Identifying a numerical spin index $\mu=+\frac12$ ($-\frac12$) with the state $\up$ ($\dn$), this drive Hamiltonian can be diagonalized in its momentum index $q$ by the gauge transformation $\c_{q\mu}^\dag\to \c_{q-\mu\phi,\mu}^\dag$ (equivalently $\c_{j\mu}^\dag\to e^{\i\phi\mu j} \c_{j\mu}^\dag$), which takes
\begin{align}
  \H_{\t{drive}}^{(\phi)} \to \H_{\t{drive}} \equiv \Omega \S_\x,
  &&
  \S_\x \equiv \sum_q \s_{\x,q},
  \label{eq:drive_trans}
\end{align}
where $\s_{\x,q}=\frac12 \c_{q,\up}^\dag \c_{q,\dn} + \t{h.c.}$ for two-level spins.

The two-level SOC drive in Eq.~\eqref{eq:drive_2} has been implemented with an external laser that couples the two electronic states of nuclear-spin-polarized atoms, with $\dn$ ($\up$) indexing the ground (excited) electronic state \cite{wall2016synthetic, livi2016synthetic, kolkowitz2016spinorbitcoupled, bromley2018dynamics, he2019engineering}.
In contrast, the drive we considered in Section \ref{sec:controls} addresses electronic excitations off-resonantly, inducing an effective Hamiltonian in the ground-state hyperfine manifold with spin projections $\mu\in\set{s,s-1,\cdots,-s}$ (a similar scheme was used to study SOC in a subspace of the ground-state manifold in Ref.~\cite{mancini2015observation}).
Nonetheless, both the two-level drive in Eq.~\eqref{eq:drive_2} and the $n$-level drive in Eq.~\eqref{eq:H_3LD} become homogeneous (i.e.~independent of the spatial mode index $j$ or $q$) and independent of the SOC angle $\phi$ after the same spin-symmetric gauge transformation\footnote{The ``asymmetric'' gauge transformation $(\c_{j,\up}^\dag,\c_{j,\dn}^\dag)\to(e^{\i\phi j} \c_{j,\up}^\dag,\c_{j,\dn}^\dag)$, sometimes performed in the two-state SOC literature, does not generalize as nicely to $n>2$.} $\c_{j\mu}^\dag\to e^{\i\phi\mu j} \c_{j\mu}^\dag$.

\begin{figure}
\centering
\includegraphics{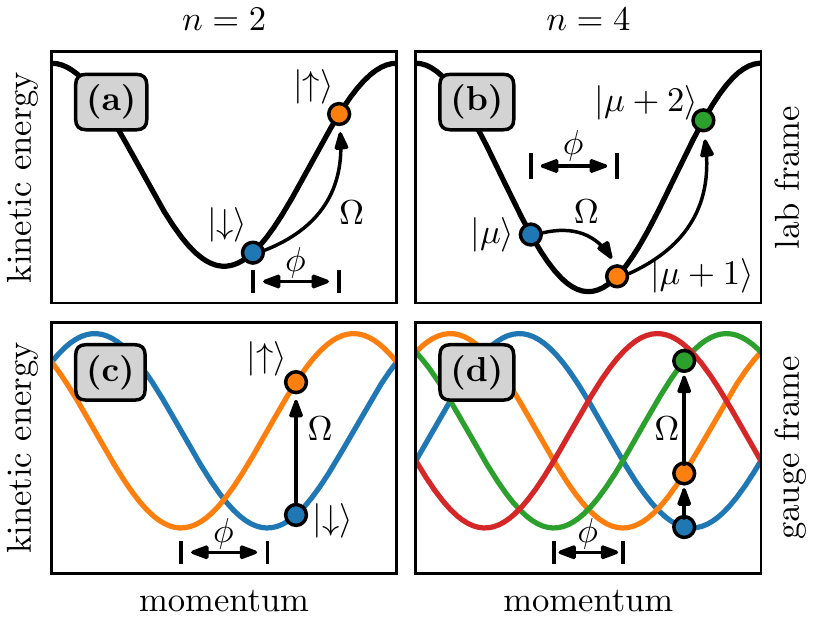}
\caption{
Spin-orbit coupling for 2-level ({\bf a},{\bf c}) and 4-level ({\bf b},{\bf d}) spins.
Colors indicate different spin projections $\mu$.
In the ``lab frame'' ({\bf a},{\bf b}), kinetic energy is insensitive to spin, but a spin transition $\mu\to\mu+1$ is accompanied by a momentum kick $q\to q+\phi$ from the drive.
Changing into the ``gauge frame'' ({\bf c},{\bf d}), essentially by shifting the momentum label $q$ for each spin state $\mu$, makes the drive diagonal in the momentum index, but comes at the cost of making kinetic energy spin-dependent.
}
\label{fig:soc_panels}
\end{figure}

Of course, spin-orbit coupling cannot be ``gauged away'' entirely.
Making a gauge transformation to simplify the drive comes at the cost of making the kinetic energy in Eq.~\eqref{H_kin_start} spin-dependent, taking
\begin{align}
  \H_{\t{kin}} \to \H_{\t{kin}}^{(\phi)}
  \equiv -2J \sum_q \cos\p{q+\mu\phi} \s_{\mu\mu q},
\end{align}
as visualized in Figure \ref{fig:soc_panels}.
To better interpret this Hamiltonian, we can write it in the form
\begin{align}
  \H_{\t{kin}}^{(\phi)}
  = -2J \sum_q
  \sp{\cos\p{q} \hat w_{+,q}^{(\phi)} - \sin\p{q} \hat w_{-,q}^{(\phi)}},
\end{align}
where
\begin{align}
  \hat w_{+,q}^{(\phi)} &\equiv \sum_\mu \cos\p{\mu\phi} \s_{\mu\mu q}, \\
  \hat w_{-,q}^{(\phi)} &\equiv \sum_\mu \sin\p{\mu\phi} \s_{\mu\mu q}.
\end{align}
For two-level spins with $\mu=\pm\frac12$, $\hat w_{+,q}^{(\phi)}$ is proportional to the identity operator and $\hat w_{-,q}^{(\phi)} = 2\sin\p{\phi/2} \s_{\z,q}$, so the kinetic Hamiltonian in the gauge frame describes a (synthetic) inhomogeneous magnetic field:
\begin{align}
  \left. \H_{\t{kin}}^{(\phi)} \right|_{n=2}
  = 4 J \sin\p{\phi/2} \sum_q \sin\p{q} \s_{\z,q}.
\end{align}
When $n>2$, an inhomogeneous magnetic field is likewise recovered in the weak SOC limit $s\phi\ll1$, in which case
\begin{align}
  \left. \H_{\t{kin}}^{(\phi)} \right|_{s\phi\ll1}
  = 2J\phi \sum_q \sin\p{q} \s_{\z,q} + O\p{(s\phi)^2}.
\end{align}
For larger $\phi$, this Hamiltonian acquires terms with higher powers of $\s_{\z,q}$, up to $\s_{\z,q}^{n-1}$.

Finally, the gauge transformation $\c_{q\mu}^\dag\to \c_{q-\mu\phi,\mu}^\dag$ also transforms the interaction Hamiltonian.
Applying this transformation to Eq.~\eqref{eq:H_int_momenta} and keeping only terms that respect coherences that can be imposed on initial states by the laser drive in Section \ref{sec:controls} (applied to an initially spin-down-polarized state) again results in an effective spin model.
For sufficiently weak SOC ($s\phi\to0$) this spin model is still well-approximated by $\H_{\t{int}}$ in Eqs.~\eqref{eq:H_int_start} and \eqref{eq:H_int}.
The validity of this approximation has been previously benchmarked for SU(2)-symmetric interactions \cite{he2019engineering, smale2019observation}, and we provide additional benchmarking for SU(4) and SU(6) in Appendix \ref{sec:benchmarking} (which finds that the spin model works well even for {\it large} $\phi$).
To ensure that $\H_{\t{kin}}^{(\phi)}$ does not become trivial as $\phi\to0$, we can keep $J\phi/u$ constant, either by increasing $J/U$ or decreasing $N/L$.
Altogether, the interacting spin Hamiltonian in the gauge frame becomes
\begin{align}
  \H_\spin = -\f{u}{2N} \v\S\cdot\v\S + 2J\phi \sum_q \sin\p{q} \s_{\z,q},
  \label{eq:H_spin}
\end{align}
consisting of a spin-locking $\v\S\cdot\v\S$ term that energetically favors permutational symmetry, and an inhomogeneous magnetic field that causes inter-spin dephasing.

\section{Mean-field theory and dynamical phases}
\label{sec:mean_field}

We now study the dynamical behavior of the SOC spin Hamiltonian $H_\spin$ in Eq.~\eqref{eq:H_spin}, and henceforth work exclusively in the ``gauge frame'' of $\H_\spin$ and the three-laser drive $\H_{\t{3LD}}$ in Eq.~\eqref{eq:H_3LD}.
We use a Ramsey-like setup wherein we prepare an initial state with the three-laser drive (using fast pule sequences), then let the state evolve freely for some time under $\H_\spin$, and finally apply again the three-laser drive to map observables of interest onto spin projection measurements (e.g.~with spin qudit tomography \cite{newton1968measurability, perlin2020qudit}).
At the mean-field (MF) level, the undriven spin Hamiltonian (neglecting constant energy shifts) becomes
\begin{align}
  \H_\MF = u \sum_q\sp{-\bk{\bar{\v s}}\cdot\v\s_q
    + h \sin\p{q} \s_{\z,q}}.
  \label{eq:H_MF}
\end{align}
where $\bar{\v s}\equiv\frac1N\sum_q\v\s_q$ is the average spin matrix, and $h\equiv 2J\phi/u$ is a dimensionless strength of the inhomogeneous magnetic field.
We assume that all momenta $q\in\ZZ_N\times 2\pi/N$ are occupied.
Fixing the atom number $N$, the spin Hamiltonian has one free parameter, $h$, which determines the relative strength of the single-particle and interaction terms.
One should therefore expect distinct dynamical behaviors when $h\ll1$, in which case strong spin-locking interactions should give rise to a long-range ordered phase, as opposed to $h\gg1$, in which case long-range order should be destroyed by the strong inhomogeneous magnetic field \cite{smale2019observation}.

To investigate these behaviors quantitatively, we examine time-averaged observables of the form
\begin{align}
  \bbk{\O}_\MF = \lim_{T\to\infty} \f1T \int_0^T \dd t \bk{\O\p{t}}_\MF,
\end{align}
where $\bk{\O\p{t}}_\MF$ is the mean-field value of observable $\O$ at time $t$.
Specifically, we consider the time-averaged magnetization
\begin{align}
  \sigma_\MF \equiv \abs{\bbk{\vec\sigma}_\MF},
  &&
  \vec\sigma \equiv \f1{Ns} \times \vec S,
\end{align}
where $\vec S\equiv(\S_\x,\S_\y,\S_\z)$ with $\S_\alpha \equiv \sum_q \s_{\alpha,q}$, and the time-averaged (dimensionless) interaction energy
\begin{align}
  \bbk{\sds}_\MF = \f1{N^2} \times \bbk{\v\S\cdot\v\S}_\MF.
\end{align}
By design, these non-negative quantities are normalized to lie on the interval $[0,1]$, independent of the system size $N$ or spin dimension $n$.
In the remainder of this section we will assume that $n$ is even, both for the sake of experimental relevance (most relevant atomic nuclei are fermionic) and to avoid complications from parity effects.

Our numerical simulations of mean-field dynamics are performed with a Schwinger boson decomposition of spin operators: $\s_{\mu\nu q} = \b_{\mu q}^\dag\b_{\nu q}$.
This decomposition requires no approximations, and reduces the number of variables to keep track of by a factor of $\sim n$.
See Appendices \ref{sec:MFT} and \ref{sec:bosons} for additional details about our numerical simulations and the Schwinger boson equations of motion.

\subsection{Initial spin-polarized state}

\begin{figure}
\centering
\includegraphics{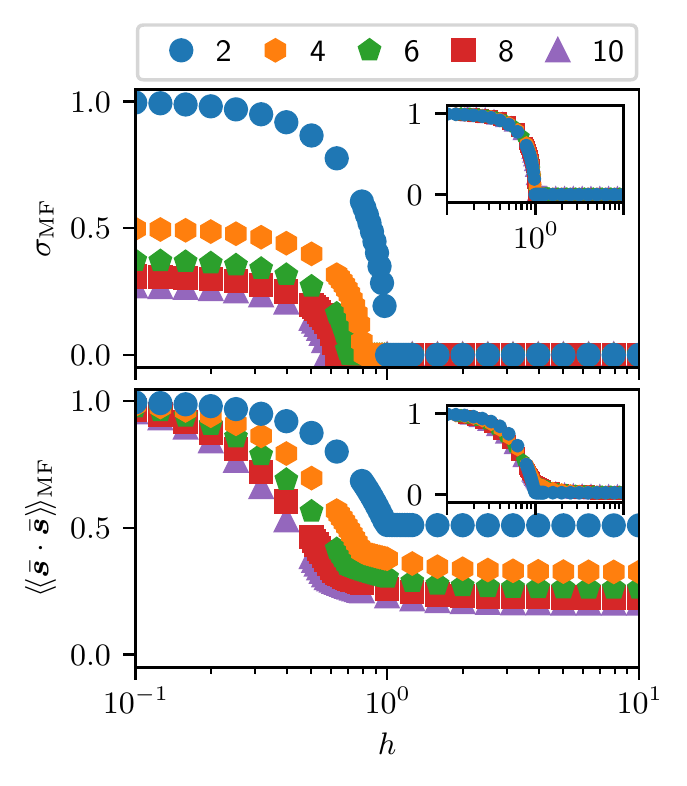}
\caption{
Time-averaged magnetization $\sigma_\MF$ and interaction energy $\bbk{\sds}_\MF$ for different spin dimensions $n$ (indicated in the legend) as determined by mean-field simulations of $N=100$ spins initially in the x-polarized state $\ket\X$ for a time $T=10^5/u$.
Insets show same data after rescaling $h\to h \times (n/2)^{1/3}$, and transforming vertical axes according to Eq.~\eqref{eq:rescale}.
}
\label{fig:mean_mag_int_X}
\end{figure}

\begin{figure}
\centering
\includegraphics{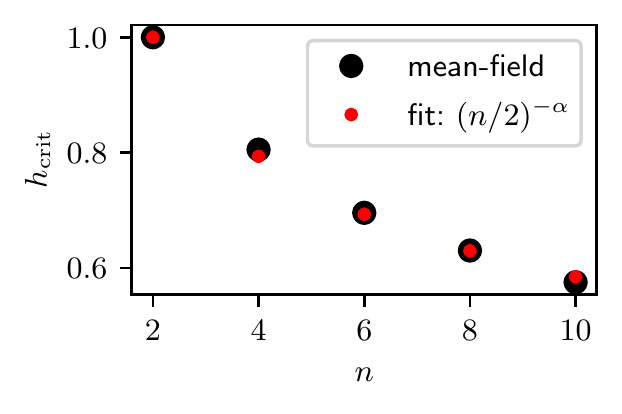}
\caption{
The critical value of $h_\crit$ as determined by mean-field simulations of $N=100$ spins initially in the x-polarized state $\ket\X$.
A single-parameter fit to $h_\crit=\p{n/2}^{-\alpha}$ finds $\alpha=0.333(5)$, and $\alpha=1/3$ is consistent with all mean-field results to within an uncertainty determined by the resolution of $h$ in mean-field simulations.
}
\label{fig:crit_fields_X}
\end{figure}

Figure \ref{fig:mean_mag_int_X} shows the time-averages of the magnetization $\sigma_\MF$ and interaction energy $\bbk{\sds}_\MF$ as computed by mean-field simulations of $N=100$ spins initially in the x-polarized state $\ket\X\equiv\ket\x^{\otimes N}$, where
\begin{align}
  \ket\x \equiv e^{-\i\frac{\pi}{2}\s_\y} \ket{s}
  = \f1{2^s} \sum_\mu { 2s \choose s+\mu }^{1/2} \ket{\mu}.
  \label{eq:state_x}
\end{align}
Here ${ m \choose k }$ is a binomial coefficient.
As expected, the spin model exhibits a mean-field dynamical phase transition between an ordered phase at small $h$ and a disordered phase at large $h$.
The ordered phase has a non-zero magnetization $\sigma_\MF$ and an interaction energy $\bbk{\sds}_\MF$ that asymptotically approach their maximal values as $h\to0$.
The disordered phase has no (time-averaged) magnetization, $\sigma_\MF=0$, but the interaction energy $\bbk{\sds}_\MF$ nonetheless indicates persistent nontrivial inter-spin correlations when $n>2$.
These nontrivial correlations vanish as $h\to\infty$, in which case $\bbk{\sds}_\MF$ approaches the minimal value allowed by conservation laws (clarified below).
By minimizing the reduced field $h$ for which $\sigma_\MF=0$, we numerically find that the transition between ordered and disordered phases occurs at a critical field $h_\crit=\p{n/2}^{-\alpha}$ with $\alpha\approx1/3$ (see Figure \ref{fig:crit_fields_X}).
When $n=2$, this transition is consistent with the predictions of a Lax vector analysis \cite{yuzbashyan2005nonequilibrium, yuzbashyan2006dynamical, yuzbashyan2006relaxation, yuzbashyan2015quantum, smale2019observation} that exploits integrability of $\H_\spin$ to determine long-time behavior.
However, additional theoretical tools are necessary to understand this transition when $n>2$.
We elaborate on this point in Appendix \ref{sec:lax}.

As shown in insets of Figure \ref{fig:mean_mag_int_X}, mean-field results for different spin dimensions $n$ collapse onto each other when normalizing the field $h$ to its critical value, $h\to h\times\p{n/2}^{1/3}$, and rescaling
\begin{align}
  \sigma_\MF \to \f{\sigma_\MF}{\gamma\p{n/2}},
  &&
  \bbk{\sds}_\MF \to \f{\bbk{\sds}_\MF-\gamma\p{n}}{1-\gamma\p{n}},
  \label{eq:rescale}
\end{align}
where
\begin{align}
  \gamma\p{k} \equiv \f{\Gamma\p{k-\frac12}}{\sqrt\pi\,\Gamma\p{k}}
  \stackrel{k\ge2}{\approx} \f1{\sqrt{\pi(k-1)}}.
  \label{eq:gamma}
\end{align}
The rescaling of magnetization and interaction energy can be understood by considering their limiting behavior as $h\to\infty$ or $h\to0$.

In the strong-field limit $h\to\infty$, we can ignore interactions and treat spins as though they simply precess at different rates.
The time-averaged transverse magnetization $\sigma_\MF$ then trivially vanishes as $h\to\infty$.
The interaction energy $\bk{\sds}_\MF = \bk{\bar{\v s}}_\MF \cdot \bk{\bar{\v s}}_\MF + O(1/N)$, meanwhile, has contributions from:
\begin{enumerate*}
\item the diagonal parts of the mean spin matrix $\bk{\bar{\v s}}_\MF$, which are conserved by inhomogeneous spin precession, and
\item the off-diagonal parts of $\bk{\bar{\v s}}_\MF$, whose oscillations average to zero when evaluating the time average in $\bbk{\sds}_\MF$.
\end{enumerate*}
Altogether, the interaction energy $\bbk{\sds}_\MF$ in the strong-field limit is determined by the time-independent diagonal part $\diag\bk{\bar{\v s}}_\MF = \diag\op{\x}$, namely
\begin{align}
  \lim_{h\to\infty} \bbk{\sds}_\MF
  = \Tr\sp{\p{\diag\op{\x}}^2}
  = \gamma\p{n}.
  \label{eq:lim_ss}
\end{align}
The same result can be obtained by computing the time-averaged interaction energy of two spins precessing at different rates.

In the weak-field limit $h\to0$, the spin-locking $\v\S\cdot\v\S$ interactions of the Hamiltonian $\H_\spin$ energetically restrict dynamics to the permutationally symmetric (PS) manifold.
To first order in $h$, the effect of the inhomogeneous field can be acquired by projecting it onto the PS manifold, which takes $\s_{\z,q}\to\frac1N\S_\z$.
The first order effect of the inhomogeneous field thus vanishes, as
\begin{align}
  \sum_q \sin\p{q} \s_{\z,q}
  \to \sum_q \sin\p{q} \times \frac1N \S_\z
  = 0.
\end{align}
At second order in $h$, the effective Hamiltonian within the PS manifold is related to the variance of the inhomogeneous field, rather than its (vanishing) average.
On a high level, the second-order effect of the inhomogeneous field within the PS manifold thus consists of permutation-symmetrized products of two spin-z operators, $\s_{\z,p}\s_{\z,q}$ (with $p,q$ possibly equal).
Altogether, the effective spin Hamiltonian at second order in $h$ is (see Appendix \ref{sec:pert_theory})
\begin{align}
  \H_\spin^\eff
  = \f{h^2 u}{2(N-1)} \times \sp{\S_\z^2 - N\sum_q \s_{\z,q}^2},
  \label{eq:H_spin_eff}
\end{align}
which in the mean-field approximation becomes
\begin{align}
  \H_\MF^\eff = -\frac12 h^2 u \sum_q \s_{\z,q}^2,
\end{align}
where we have used the fact that the axial magnetizations $\bk{\s_{\z,q}}=\frac1N\bk{\S_\z}$ within the PS manifold, and the initial value of $\bk{\S_\z}=0$ is conserved by $\H_\spin$.
The weak-field effective Hamiltonian preserves permutational symmetry, so $\bbk{\sds}_\MF\to1$ as $h\to0$.
Moreover, the initial y-magnetization $\bk{\S_\y}=0$ is conserved by $\H_\spin$, so the long-time-averaged magnetization $\sigma_\MF$ is determined by the time-average of $\s_\x$ for a single (any) spin:
\begin{align}
  \lim_{h\to0} \sigma_\MF
  = \f1s \abs{\lim_{T\to\infty} \f1T \int_0^T \dd\tau
  \bk{\x|\s_\x\p{\tau}|\x}},
\end{align}
where
\begin{align}
  \s_\x\p{\tau} = e^{\i\tau \s_\z^2} \s_\x e^{-\i\tau \s_\z^2}.
\end{align}
We can adapt exact analytical results for the dynamics of an infinite-range Ising model \cite{foss-feig2013nonequilibrium}\footnote{See Appendix K of Ref.~\cite{perlin2020shorttime} for a simpler adaptation of the analytics in Ref.~\cite{foss-feig2013nonequilibrium} to the one-axis twisting model $\H_{\t{OAT}}=\chi \s_\z^2$.} to find that
\begin{align}
  \bk{\x|\s_\x\p{\tau}|\x} = s \p{\cos\tau}^{n-2},
\end{align}
so for even $n$
\begin{align}
  \lim_{h\to0} \sigma_\MF
  = \f1{2\pi} \int_0^{2\pi} \dd\tau \p{\cos\tau}^{n-2}
  = \gamma\p{\f{n}{2}}.
\end{align}
When going beyond mean-field theory, inter-spin correlations generated by $\S_\z^2$ in Eq.~\eqref{eq:H_spin_eff} will cause $\bk{\S_\x}$ (and thereby the magnetization $\bk{\vec\sigma}$) to decay as $e^{-O(t^2/Ns)}$; the timescale of this decay diverges as $N\to\infty$.
On a lattice of linear size $L$ without periodic boundary conditions, additional corrections to the behavior predicted above will appear on $O(L/J)$ timescales.

\subsection{Initial kitten states}

\begin{figure}
\centering
\includegraphics{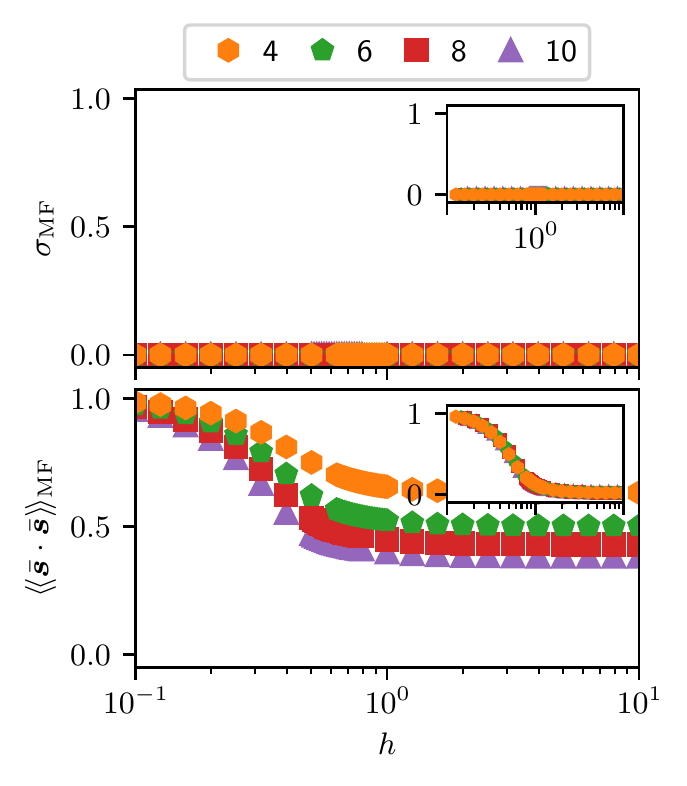}
\caption{
A corollary of Figure \ref{fig:mean_mag_int_X} for the initial state $\ket\XX$.
The inset for interaction energy $\bbk{\sds}_\MF$ in Figure \ref{fig:mean_mag_int_X} subtracts off the minimal value of $\bbk{\sds}_\MF$ and rescales to lie on the interval $[0,1]$, as prescribed in Eq.~\eqref{eq:rescale}.
Here the subtracting and rescaling is identical, but with a minimal value of $2\gamma\p{n}$ rather than $\gamma\p{n}$.
}
\label{fig:mean_mag_int_XX}
\end{figure}

\begin{figure}
\centering
\includegraphics{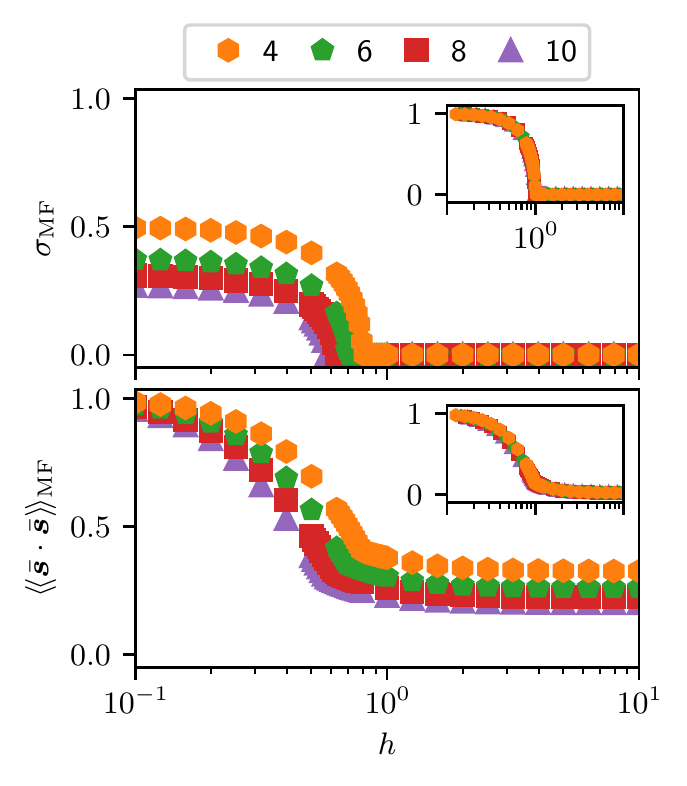}
\caption{
A corollary of Figure \ref{fig:mean_mag_int_X} for the initial state $\ket\XXI$.
Insets show the same data shifted and rescaled identically to Figure \ref{fig:mean_mag_int_X}.
}
\label{fig:mean_mag_int_XXI}
\end{figure}

We now consider the same setup as above, but with the initial ``kitten'' states $\ket\XX\equiv\ket\xx^{\otimes N}$ and $\ket\XXI\equiv\ket\xxi^{\otimes N}$, where
\begin{align}
  \ket\xx \equiv \f{\ket{\x} + \ket{-\x}}{\sqrt{2}},
  &&
  \ket\xxi \equiv \f{\ket{\x} + \p{-1}^s \ket{-\x}}{\sqrt{2}},
\end{align}
and $\ket{-\x}$ is a state polarized along $-x$, defined similarly to $\ket\x$ in Eq.~\eqref{eq:state_x}:
\begin{align}
  \ket{-\x} \equiv e^{-\i\frac{\pi}{2}\s_\y}\ket{-s}
  = \f1{2^s} \sum_\mu \p{-1}^{s+\mu}
  { 2s \choose s+\mu }^{1/2} \ket{\mu}.
\end{align}
Similarly to Figure \ref{fig:mean_mag_int_X}, Figures \ref{fig:mean_mag_int_XX} and \ref{fig:mean_mag_int_XXI} show the time-averaged magnetization $\sigma_\MF$ and interaction energy \mbox{$\bbk{\sds}_\MF$} throughout mean-field dynamics of the initial states $\ket\XX$ and $\ket\XXI$.
These figures exclude the trivial case of spin dimension $n=2$, for which $\ket\xx=\ket{-s}$ is an eigenstate of $\H_\spin$ and $\ket\xxi=e^{-\i\frac{\pi}{2}\s_\z}\ket\x\equiv\ket\y$ is spin-polarized along the $y$ axis.
The first and perhaps most interesting observation to make about Figures \ref{fig:mean_mag_int_XX} and \ref{fig:mean_mag_int_XXI} is that they are different, signifying the importance of intra-spin coherences for the dynamical behavior of multilevel spin models.

Unlike Figure \ref{fig:mean_mag_int_X} (for $\ket\X$), Figure \ref{fig:mean_mag_int_XX} (for $\ket\XX$) exhibits no sharp transition between distinct dynamical phases: the time-averaged magnetization $\sigma_\MF=0$ for all values of the field $h$, and the interaction energy $\bbk{\sds}_\MF$ smoothly crosses over from a maximal value of 1 to a minimal value of $2\gamma\p{n}$.
The minimal value of $\bbk{\sds}_\MF$ approached as $h\to\infty$ can be explained with arguments identical to those in the paragraph containing Eq.~\eqref{eq:lim_ss}, which now imply that
\begin{align}
  \lim_{h\to\infty} \bbk{\sds}_\MF
  = \Tr\sp{\p{\diag\op{\xx}}^2}
  = 2\gamma\p{n}.
\end{align}
The vanishing magnetization $\sigma_\MF=0$ in Figure \ref{fig:mean_mag_int_XX} is protected by symmetries of $\H_\spin$ and $\ket\XX$.
For all initial states that we have considered, the value of $\bk{\S_\z}=0$ is conserved by the spin Hamiltonian $\H_\spin$.
Moreover, both the spin Hamiltonian $\H_\spin$ and the state $\ket\XX$ are invariant (up to global phase) under the action of $\R_\z^\pi$, where $\R_\z^\theta\equiv e^{-\i\theta \S_\z}$, which is to say that
\begin{align}
  \R_\z^\pi \H_\spin \R_\z^\pi{}^\dag = \H_\spin
  &&
  \R_\z^\pi \ket\XX \simeq \ket\XX,
\end{align}
where $\simeq$ denotes equality up to an overall phase.
This symmetry implies that
\begin{align}
  &\bk{\S_\x} = \bk{\R_\z^\pi{}^\dag \S_\x \R_\z^\pi} = -\bk{\S_\x} = 0, \\
  &\bk{\S_\y} = \bk{\R_\z^\pi{}^\dag \S_\y \R_\z^\pi} = -\bk{\S_\y} = 0
\end{align}
at all times, so altogether $\sigma_\MF=0$.

Turning now to mean-field results for the initial kitten state $\ket\XXI$ in Figure \ref{fig:mean_mag_int_XXI}, we remark that the magnetization $\sigma_\MF$ and interaction energy $\bbk{\sds}_\MF$ behave identically to those for the initial spin-polarized state $\ket\X$ in Figure \ref{fig:mean_mag_int_X}.
This finding can be understood through the fact that
\begin{align}
  \ket\XXI \simeq \R_\z^{\pi/2} \T_\z^{\pi/2} \ket\X,
\end{align}
where $\T_\z^\theta\equiv e^{-\i\theta \S_\z^2}$.
The operators $\R_\z^\theta$ and $\T_\z^\theta$ are generated by axial fields that respect permutational symmetry, and therefore commute with the spin Hamiltonian $\H_\spin$, so
\begin{align}
  e^{-\i t \H_\spin} \ket\XXI
  &\simeq e^{-\i t \H_\spin} \R_\z^{\pi/2} \T_\z^{\pi/2} \ket\X \\
  &\simeq \R_\z^{\pi/2} \T_\z^{\pi/2} e^{-\i t \H_\spin} \ket\X.
\end{align}
In turn, expanding $\sds$ according to Eq.~\eqref{eq:dot_product} shows that
\begin{align}
  \T_\z^\theta{}^\dag \R_\z^\theta{}^\dag \, \sds \, \R_\z^\theta \T_\z^\theta
  = \sds,
\end{align}
which implies that the interaction energy $\bk{\sds}$ throughout dynamics of the initial kitten state $\ket\XXI$ is the same as that of the spin-polarized state $\ket\X$.

To make sense of why the magnetization $\sigma_\MF$ is identical in Figure \ref{fig:mean_mag_int_XXI} for $\ket\XXI$ as in Figure \ref{fig:mean_mag_int_X} for $\ket\X$, we follow a four-part argument:
\begin{enumerate}
\item \label{pt:spin_mat} The time-averaged magnetization vector $\bbk{\vec\sigma}_\MF$ can be written as a function of the time-averaged spin matrix $\bbk{\bar{\v s}}_\MF$.
\item \label{pt:diags} The spin matrix $\bbk{\bar{\v s}}_\MF$ is only ever nonzero on its diagonal and anti-diagonal, regardless of the initial state.
That is, nonzero components $\bbk{\bar s_{\mu\nu}}_\MF$ of $\bbk{\bar{\v s}}_\MF$ always have $\mu=\pm\nu$ (see discussion below).
\item \label{pt:twist} The twist operator $\T_\z^\theta$ acts trivially on the diagonal and anti-diagonal components of $\bar{\v s}$, which together with point \ref{pt:diags} implies that $\bbk{\T_\z^\theta{}^\dag \, \bar{\v s} \, \T_\z^\theta}_\MF = \bbk{\bar{\v s}}_\MF$.
\item \label{pt:rot} The rotation operator $\R_\z^\theta$ merely rotates the magnetization vector $\bbk{\vec\sigma}_\MF$ without changing its magnitude.
\end{enumerate}
Altogether, points \ref{pt:spin_mat}--\ref{pt:rot} imply that the magnetization
\begin{align}
  \sigma_\MF = \abs{\bbk{\vec\sigma}_\MF}
  = \abs{\bbk{\T_\z^\theta{}^\dag \R_\z^\theta{}^\dag \,
      \vec\sigma \, \R_\z^\theta \T_\z^\theta}_\MF}
\end{align}
is the same for the initial state $\ket\XXI$ as for $\ket\X$.

The only nontrivial step in the above argument is point \ref{pt:diags}, which says that $\bbk{\bar s_{\mu\nu}}_\MF$ is guaranteed to be zero unless $\mu=\pm\nu$.
This observation, nominally a numerical result of mean-field simulations, can be understood as follows.
The eigenstates $\ket{m,w}$ of $\H_\spin$ are uniquely identified by definite numbers $m=(m_s,m_{s-1},\cdots,m_{-s})$ of atoms occupying each internal spin state $\mu\in\set{s,s-1,\cdots,-s}$, and an auxiliary index $w$ that encodes how $\ket{m,w}$ transforms under permutations of all spins (see Appendix \ref{sec:pert_theory})\footnote{Seen otherwise, since $\S_{\mu\mu}$ commutes with $\H_\spin$, eigenvectors of $\H_\spin$ can be indexed by eigenvalues of $\S_{\mu\mu}$.
The number $m_\mu$ is then the eigenvalue of $\ket{m,w}$ with respect to $\S_{\mu\mu}$, i.e.~$\S_{\mu\mu}\ket{m,w} = m_\mu\ket{m,w}$, while $w$ encodes all other information required to uniquely specify $\ket{m,w}$.}.
The operator $\bar s_{\mu\nu}=\frac1N \S_{\mu\nu}$ with $\mu\ne\nu$ couples the state $\ket{m,w}$ to states $\ket{m',w'}$ in which $(m_\mu',m'_\nu)=(m_\mu+1,m_\nu-1)$.
Generically, states $\ket{m,w}$ and $\ket{m',w'}$ with $m\ne m'$ will have different energies, so their coherence oscillates and averages to zero when evaluating time-averaged expectation values.

However, degeneracies yield stationary (time-independent) coherences that survive time-averaging.
In the weak-field limit $h\to0$, such a degeneracy occurs at the mean-field level between PS states differing only in the populations $m_\mu,m_{-\mu}$ (with a fixed value of $m_\mu+m_{-\mu}$), as the effective Hamiltonian becomes $\H_\MF^\eff\propto\sum_\mu\mu^2m_\mu$.
This symmetry is preserved at all orders in perturbation theory\footnote{Only even powers of the ``perturbation'' $\sum_q\sin\p{q}\s_{\z,q}$ can be nonzero within the PS manifold, and even powers of this perturbation exhibit the same mean-field degeneracy between states differing only in the populations $m_\mu,m_{-\mu}$.}, so some coherence between such states is preserved as $h\to h_\crit$, although this coherence decays as perturbative corrections to degenerate eigenstates cause them to leak out of the PS manifold (and thereby have a smaller overlap with the initial state $\ket\X$).
Note that beyond-mean-field effects break the symmetry protecting anti-diagonal components of $\bbk{\v\s}_\MF$, causing them to decay on time scales that should diverge as $N\to\infty$.

\section{Conclusions and future directions}
\label{sec:conclusions}

Starting with an SU($n$) Hubbard model describing ultracold fermionic alkaline-earth(-like) atoms on an optical lattice, we derived a momentum-space multilevel spin model with all-to-all SU($n$)-symmetric interactions.
We then introduced external control fields, finding a simple three-laser drive that homogeneously addresses nuclear spins with a variety of spin Hamiltonians.
Taking a closer look at the effect of the spin-orbit coupling (SOC) induced by the driving lasers, we found that maintaining the validity of the spin model requires weak SOC, which in turn gives rise to a (synthetic) inhomogeneous magnetic field.
Finally, we examined dynamical behavior of the SU($n$) spin model at the mean-field level, finding that long-time observables obey simple scaling relations with $n$, and that when $n>2$ dynamical behavior can be highly sensitive to intra-spin coherences.

Our work makes important progress in understanding the SU($n$) Fermi-Hubbard model in experimentally relevant parameter regimes, and we expect our findings to be readily testable in experiments with ultracold atoms.
Given the possibility for long-range SU($n$) interactions, we hope our work stimulates further efforts into simulating SY and SYK-like models \cite{sachdev1993gapless, bentsen2019integrable} in cold atomic platforms.
In follow-up work, it would be interesting to study the relationship between initial states and dynamical phases of our SU($n$) spin model more systematically, and to consider the effect of quantum corrections to mean-field behavior.
There is also room to improve on the three-laser drive introduced in this work, for which it is natural to ask what additional techniques or ingredients are necessary to implement universal control of individual nuclear spins.
Universal control would allow for an experimental study of $n$-dependence (including even/odd-$n$ parity effects) in a single experimental platform, simply by controlling the occupation and coherence of internal spin states.
Finally, one can also study the SU($n$) Hubbard model in the super-exchange regime that gives rise to a real-space (as opposed to momentum-space) spin model, where SOC gives rise to chiral multilevel spin interactions.
Unlike our present work, the super-exchange regime does not require weak SOC, and therefore has a larger parameter space in which to explore dynamical behavior.

\section*{Acknowledgements}

We thank Victor Gurarie, Emil Yuzbashyan, Asier P.~Orioli, and Jeremy T.~Young for helpful discussions on this work.
This work was supported by the AFOSR grants FA9550-18-1-0319, FA9550-19-1-027, by the DARPA and ARO grant W911NF-16-1-0576, ARO W911NF-19-1-0210, DOE (QSA), NSF PHY1820885, NSF JILA-PFC PHY-1734006, NSF QLCI-2016244, and by NIST.

\bibliography{main.bib}
\onecolumngrid
\appendix

\section{Numerical benchmarking of the spin model}
\label{sec:benchmarking}

\begin{figure}[!h]
\centering
\includegraphics[width=\textwidth]{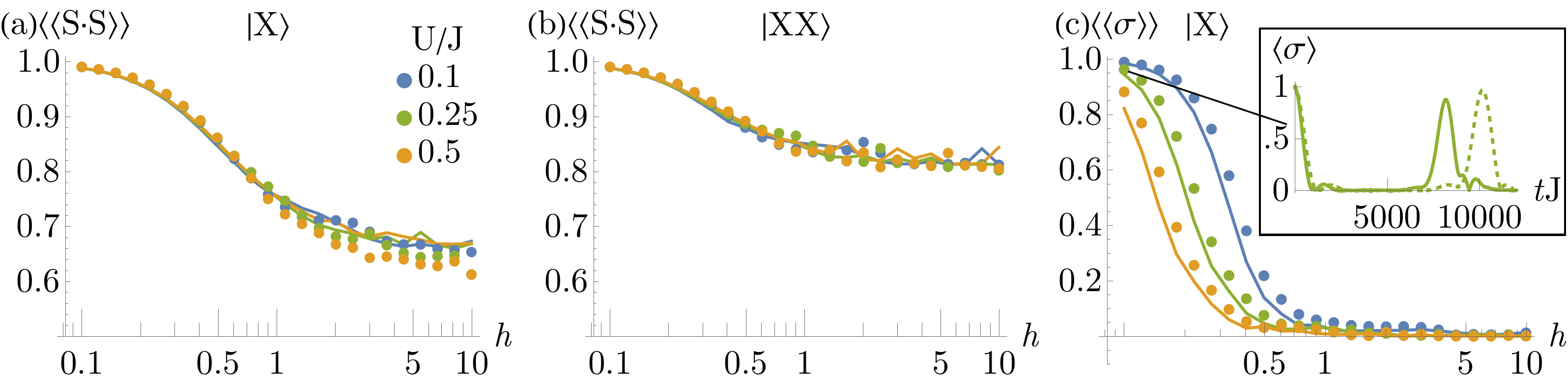}
\caption{
Numerical results (analogous to Figures \ref{fig:mean_mag_int_X}, \ref{fig:mean_mag_int_XX}, and \ref{fig:mean_mag_int_XXI} of the main text) for the time-averaged interaction energy and magnetization (both normalized to a maximal value of 1) in a system of $L=5$ lattice sites, for both a Fermi-Hubbard model (dots) and spin model (lines) with $n=4$ internal states per spin.
The corresponding initial state (defined in Section \ref{sec:mean_field} of the main text) is indicated in each panel, and observables are averaged over a time $tJ=200$.
Color indicates the value of $U/J$, and the field $h$ corresponds to $2J\phi/u$ in the case of the Fermi-Hubbard model.
Simulations are performed in real-space, with spin-orbit coupling (SOC) implemented through a homogeneous drive (with no site or $\phi$ dependence) and nearest-neighbor tunneling terms that contain factors of $e^{\pm\i\mu\phi}$.
Results for the initial kitten state $\ket\XXI$ are excluded because they are identical to those of $\ket\X$, and magnetization for the initial state $\ket\XX$ is always $0$.
Note that while panels (a) and (b) are representative of infinite-time behavior, the inset in panel (c) shows that the Fermi-Hubbard and spin models exhibit different behaviors on very long time scales, although good agreement is restored by rescaling time in the spin model, indicating the likelihood of a need to renormalize spin model parameters.
In any case, such time scales are inaccessible in current experiments and diverge as $N\to\infty$, so these corrections do not affect the main results of our work.
}
\label{fig:benchmarking_SU4}
\end{figure}

\begin{figure}[!h]
\centering
\includegraphics[width=\textwidth]{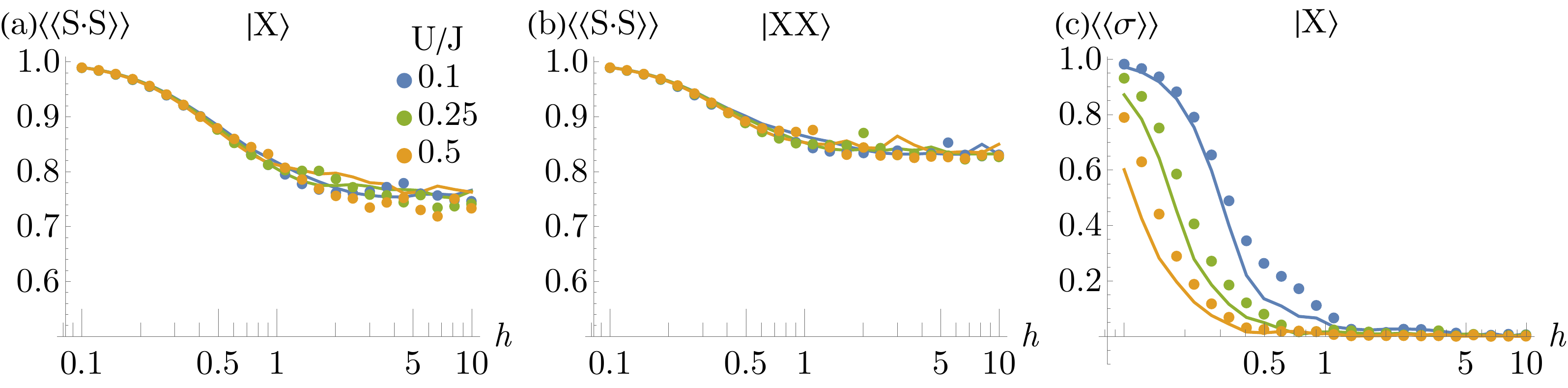}
\caption{
Numerical results identical to Figure \ref{fig:benchmarking_SU4}, but with $L=4$ lattice sites and $n=6$ internal states per spin.
}
\label{fig:benchmarking_SU6}
\end{figure}

In this appendix we present numerical evidence to support the validity of the spin models derived in Sections \ref{sec:spin_model} and \ref{sec:SOC}.
Figures \ref{fig:benchmarking_SU4} and \ref{fig:benchmarking_SU6} show a set of time-averaged observables computed with numerically exact simulations of a Fermi-Hubbard model and an effective spin model, respectively, with $n=4$ (Figure \ref{fig:benchmarking_SU4}) and $n=6$ (Figure \ref{fig:benchmarking_SU6}) internal levels per spin.
Details for these simulations are provided in the caption of Figure \ref{fig:benchmarking_SU4}.
Our main conclusion from these figures is that the two models show remarkable agreement for the observables considered in our work.
Note that these results are only intended to benchmark the approximation of a Fermi-Hubbard model by a spin model; these results are not expected to agree with the mean-field theory in Section \ref{sec:mean_field} due to strong finite-size effects.

\section{Perturbation theory for SU($n$) ferromagnets}
\label{sec:pert_theory}

Here we work out a general perturbation theory for SU($n$) ferromagnets with a gapped permutationally symmetric (PS) manifold.
We begin with an SU($n$)-symmetric interaction Hamiltonian of the form
\begin{align}
  \H_0 = \sum_{i<j} g_{ij} \, \hat\Pi_{ij},
  &&
  \hat\Pi_{ij} \equiv \v\s_i\cdot\v\s_j
  = \sum_{\mu,\nu} \s_{\mu\nu i} \s_{\nu\mu j},
  \label{eq:H_0}
\end{align}
where $g_{ij}$ are (real) scalar coefficients for the permutation operators $\hat\Pi_{ij}$, and $\s_{\mu\nu i}\equiv \c_{\mu i}^\dag \c_{\nu i}$ is a transition operator for spin $i$.
We can then consider the addition of, for example, an inhomogeneous magnetic field or Ising couplings,
\begin{align}
  \H_{\t{field}} = \sum_i B_i \s_{\z,i},
  &&
  \H_{\t{Ising}} = \sum_{i\ne j} J_{ij} \s_{\z,i} \s_{\z,j},
  \label{eq:field_ising}
\end{align}
or more generally an $M$-body operator\footnote{At face value, an $M$-body operator with $M>2$ does not typically appear in experiments.
Nonetheless, considering $M>2$ illuminates the structure of eigenstates (and eigenvalues) of $\H_0$, and allows us to go to high orders in perturbation theory with single- and two-body perturbations.}
\begin{align}
  \O(w,\hat X) = \sum_{k\in\D_N\p{M}} w_k \hat X_k,
\end{align}
where $w$ is a dimension-$M$ (i.e.~$M$-index) tensor of scalar
coefficients $w_k\equiv w_{k_1k_2\cdots k_M}$; $X$ is an $M$-spin
operator, e.g.~$\s_\z\otimes \s_\z$ in the case of Ising interactions
with $M=2$; $k\equiv\p{k_1,k_2,\cdots,k_M}$ is a list of the
individual spins $k_i\in\ZZ_N\equiv\set{1,2,\cdots,N}$ that the operator $\hat X_k\equiv \hat X_{k_1k_2\cdots k_M}$ acts on; and
\begin{align}
  \D_N\p{M} \equiv
  \set{ k \in \ZZ_N^M : \t{all entries $k_i$ of $k$ are distinct} },
\end{align}
is the strictly ``off-diagonal'' part of $\ZZ_N^M$, which is necessary to identify for a consistent definition of $\hat X_k$ as an $M$-body operator.
In this notation, the magnetic field and Ising Hamiltonians in Eq.~\eqref{eq:field_ising} respectively become $\O\p{B,\s_{\z}}$ and $\O\p{J,\s_\z\otimes \s_\z}$.

If the addition $\O(w,\hat X)$ to the SU($n$)-symmetric Hamiltonian $\H_0$ in Eq.~\eqref{eq:H_0} is sufficiently small, namely with operator norm $\norm*{\O(w,\hat X)}$ less than half the spectral gap $\Delta_{\t{gap}}$ of $\H_0$, $\norm*{\O(w,\hat X)}<\Delta_{\t{gap}}/2$, then we can treat the effect of $\O(w,\hat X)$ on the ground-state PS manifold $\E_0$ perturbatively.
The effective Hamiltonians $\H_\eff^{(1)}$ and $\H_\eff^{(2)}$ induced by $\O(w,\hat X)$ on the PS manifold $\E_0$ at leading orders in perturbation theory are \cite{bravyi2011schrieffer}
\begin{align}
  \H_\eff^{(1)} = \P_0 \O(w,\hat X) \P_0,
  &&
  \H_\eff^{(2)} = - \sum_{\Delta\ne0}
  \f1\Delta \P_0 \O(w,\hat X) \P_\Delta \O(w,\hat X) \P_0,
  \label{eq:H_eff}
\end{align}
where $\P_\Delta$ is a projector onto the eigenspace $\E_\Delta$ of $\H_0$ with interaction energy $\Delta$ above that of the PS manifold.
The first order effective Hamiltonian $\H_\eff^{(1)}$ simply projects $\O(w,\hat X)$ onto the PS manifold $\E_0$, and takes the form
\begin{align}
  \H_\eff^{(1)} = \mean w \, \col X,
  \label{eq:H_eff_1}
\end{align}
where the coefficient $\mean w$ is the average of all coefficients $w_k$; and $\col X$ is a collective version of $X$:
\begin{align}
  \mean w \equiv \f1{\abs{\D_N\p{M}}}
  \sum_{k\in\D_N\p{M}} w_k,
  &&
  \col X \equiv \sum_{k\in\D_N\p{M}} \hat X_k,
  \label{eq:mean_col}
\end{align}
with $\abs{\D_N\p{M}}=\prod_{j=0}^{M-1}\p{N-j}$.
In the case of a magnetic field $\s_\z$ or Ising interactions  $\s_\z\otimes \s_\z$, for example,
\begin{align}
  \col{\s_\z} = \sum_i \s_\z^{(i)} = \S_\z,
  &&
  \col{\s_\z\otimes \s_\z}
  = \sum_{i\ne j} \s_\z^{(i)} \s_\z^{(j)}
  = \S_\z^2 - N \sum_i \s_{\z,i}^2.
\end{align}
The second order effective Hamiltonian $\H_\eff^{(2)}$ in Eq.~\eqref{eq:H_eff} takes more work to simplify due to the presence of a projector $\P_\Delta$ onto the manifold $\E_\Delta$ of states with excitation energy $\Delta$.
This projector essentially picks off the part of $\O(w,\hat X)$ that is strictly off-diagonal with respect to the ground- and excited-state manifolds $\E_0$ and $\E_\Delta$.
We therefore need to decompose $\O(w,\hat X)$ into components that generate states of definite excitation energy when acting on PS states $\ket\psi\in\E_0$.
The SU($n$) symmetry of $\H_0$ enables such a decomposition to take the form
\begin{align}
  \H_0 \O(w,\hat X) \ket\psi
  = \sum_\Delta \p{E_0+\Delta} \O(w^\Delta,\hat X) \ket\psi,
  &&
  E_0 \equiv \sum_{i<j} g_{ij},
  \label{eq:excitation}
\end{align}
where $E_0$ is the interaction energy of PS states, and thinking of the tensor $w$ as a $\abs{\D_N\p{M}}$-component vector, the tensor $w^\Delta$ can be found by
\begin{enumerate*}
\item using the coefficients $g_{ij}$ to construct a matrix $g^{(M)}$ of dimensions $\abs{\D_N\p{M}}\times\abs{\D_N\p{M}}\sim N^M\times N^M$, and
\item projecting $w$ onto the eigenspace of $g^{(M)}$ with eigenvalue $\Delta$.
\end{enumerate*}
We construct $g^{(M)}$ for the single-body ($M=1$) case below (in Appendix \ref{sec:eigenstates}), and provide explicit forms of $g^{(M)}$ with arbitrary $M$.

Equipped with the decomposition $\O(w,\hat X)=\sum_\Delta\O(w^\Delta,\hat X)$ with terms $\O(w^\Delta,\hat X)$ that generate states of definite excitation energy $\Delta$, we can expand
\begin{align}
  \H_{\t{eff}}^{(2)} = -\sum_{\Delta\ne0} \f1\Delta
  \P_0 \O(w^\Delta,\hat X)^2 \P_0.
\end{align}
If $X$ is a single-body operator, then
\begin{align}
  \H_{\t{eff}}^{(2)}
  = \sum_{\Delta\ne0} \f{w^\Delta\cdot w^\Delta}{N\p{N-1}\Delta}
  \p{\col{X}^2 - N \col{X^2}},
\end{align}
and if furthermore all $g_{ij}=-U/N$, as for $\H_{\t{int}}$ in Eq.~\eqref{eq:H_int}, then the only relevant excitation energy is $\Delta=U$ (see Section \ref{sec:spin_wave}), and
\begin{align}
  w^U\cdot w^U = \sum_i \p{w_i-\mean w}^2 = N \widetilde{w}^2
\end{align}
is simply $N$ times the variance $\widetilde{w}^2$ of $w$, so
\begin{align}
  \H_{\t{eff}}^{(2)}
  = \f{\widetilde{w}^2}{\p{N-1}U} \p{\col{X}^2 - N \col{X^2}}.
\end{align}

\subsection{Generating excitation energy eigenstates}
\label{sec:eigenstates}

Here we construct the matrix $g^{(M)}$ that enables decomposing $M$-body operators $\O(w,\hat X)$ into terms $\O(w^\Delta,\hat X)$ that generate states of definite excitation energy $\Delta$ above the PS manifold, as in Eq.~\eqref{eq:excitation}.
We work through the calculation of $g^{(1)}$ explicitly, and provide the result for $g^{(M)}$ from a generalized version of the same calculation.
To this end, we consider the action of a single-body operator $\O(w,\hat X)=\sum_iw_i\hat X_i$ on an arbitrary PS state $\ket\psi\in\E_0$ and expand
\begin{align}
  \H_0 \O(w,\hat X) \ket\psi
  = \f12 \sum_{i\ne j} \sum_k g_{ij} w_k \hat\Pi_{ij} \hat X_k \ket\psi,
  \label{eq:single_body_diagnosis_start}
\end{align}
where strictly speaking $g_{ij}$ has only been defined for $i<j$, so for completeness we define $g_{ji}=g_{ij}$ and $g_{ii}=0$.
The sum in Eq.~\eqref{eq:single_body_diagnosis_start} has terms with $k\in\set{i,j}$ and terms with $k\notin\set{i,j}$.
In the case of $k\notin\set{i,j}$, the permutation operator $\hat\Pi_{ij}$ commutes with $\hat X_k$ and annihilates on $\ket\psi$, and we can replace the sum
\begin{align}
  \sum_{k\notin\set{i,j}} \to \sum_k - \sum_{k\in\set{i,j}},
\end{align}
allowing us to simplify
\begin{align}
  \f12 \sum_{i\ne j} \sum_{k\notin\set{i,j}}
  g_{ij} w_k \hat\Pi_{ij} \hat X_k \ket\psi
  = E_0 \O(w,\hat X) \ket\psi
  - \f12 \sum_{i\ne j} \sum_{k\in\set{i,j}} g_{ij} w_k \hat X_k \ket\psi,
\end{align}
where $E_0=\frac12\sum_{i\ne j}g_{ij}$ is the interaction energy the PS state $\ket\psi\in\E_0$.
Switching the order of sums over $i\ne j$ and $k\in\set{i,j}$ as
\begin{align}
  \sum_{i\ne j} \sum_{k\in\set{i,j}}
  \to \sum_k \sum_{\substack{i\ne j\\\set{i,j}\ni k}},
\end{align}
we can simplify
\begin{align}
  \f12 \sum_{\substack{i\ne j\\\set{i,j}\ni k}} g_{ij}
  = \f12 \sum_i g_{ik} + \f12 \sum_j g_{kj}
  = g_k,
  &&
  g_k \equiv \sum_i g_{ik},
\end{align}
which implies that the terms in Eq.~\eqref{eq:single_body_diagnosis_start}
with $k\notin\set{i,j}$ are
\begin{align}
  \f12 \sum_{i\ne j} \sum_{k\notin\set{i,j}}
  g_{ij} w_k \hat\Pi_{ij} \hat X_k \ket\psi
  = E_0 \O(w,\hat X) \ket\psi - \sum_k g_k w_k \hat X_k \ket\psi.
\end{align}
The terms in Eq.~\eqref{eq:single_body_diagnosis_start} with
$k\in\set{i,j}$, meanwhile, are
\begin{align}
  \f12 \sum_{\substack{i\ne j\\k\in\set{i,j}}}
  g_{ij} w_k \hat\Pi_{ij} \hat X_k \ket\psi
  = \sum_{i,j} g_{ij} w_j \hat X_i \ket\psi.
\end{align}
so in total
\begin{align}
  \H_0 \O(w,\hat X) \ket\psi
  = E_0 \O(w,\hat X) \ket\psi
  + \sum_k \sp{\sum_j g_{kj} w_j - g_k w_k} \hat X_k \ket\psi.
\end{align}
The action of the single-body perturbation $\O(w,\hat X)$ on a permutationally symmetric state therefore generates an eigenstate of $\H_0$ with interaction energy $E_0+\Delta$ if the vector $w=\sum_k w_k \ket{k}$ satisfies the eigenvalue equation
\begin{align}
  g^{(1)} \cdot w = \Delta w, && g^{(1)} \equiv g - \diag\vec g,
  \label{eq:single_body_eig}
\end{align}
where $g \equiv\sum_{i,j} g_{ij} \op{i}{j}$ is a matrix of all couplings $g_{ij}$; the vector $\vec g \equiv \sum_{i,j} g_{ij} \ket{i} = \sum_i g_i \ket{i}$ is the sum of all columns of $g$; and the matrix $\diag\vec g \equiv \sum_i g_i \op{i}$ has $\vec g$ on the diagonal and zeroes everywhere else.

A similar calculation as above with arbitrary $M$ yields an eigenvalue equation of the form
\begin{align}
  g^{(M)} \cdot w = \Delta w,
  \label{eq:multi_body_problem}
\end{align}
where we treat $w$ as an $\abs{\D_N\p{M}}$-component vector, and $g^{(M)}$ is a matrix with dimensions $\abs{\D_N\p{M}}\times\abs{\D_N\p{M}}$.
In the case of $M=2$, we have
\begin{align}
  g^{(2)} = \sum_{\p{k,\ell}\in\D_N\p{2}} \ket{k\ell}
  \sp{\sum_{\substack{i\in\ZZ_N\\i\notin\set{k,\ell}}}
    \p{ g_{ik} \bra{i\ell} + g_{i\ell} \bra{ki} }
    + g_{k\ell} \bra{\ell k}
    - \p{g_k + g_\ell - g_{k\ell}} \bra{k\ell}},
  \label{eq:two_body_mat}
\end{align}
and more generally
\begin{align}
  g^{(M)} = \sum_{k\in\D_N\p{M}} \ket{k}
  \sp{\sum_{a\in\ZZ_M} \sum_{\substack{i\in\ZZ_N\\i\notin k}}
    g_{ik_a} \bra{k_{a:i}}
    + \sum_{\set{a,b}\in\C_M\p{2}} g_{k_ak_b} \bra{k_{a\leftrightarrow b}}
    - \tilde g_k \bra{k}},
  \label{eq:multi_body_mat}
\end{align}
where $k_a\in k=\p{k_1,k_2,\cdots,k_M}$; $k_{a:i}$ a list that is equal to $k$ except at the $a$-th position, where $k_a$ replaced is by $i$, i.e.~$k_{a:i}=\p{\cdots,k_{a-1},i,k_{a+1},\cdots}$; $\C_L\p{p}$ is the set of all subsets (``choices'') of $p$ elements from $\ZZ_L$; $k_{a\leftrightarrow b}$ is equal to $k$ except at the $a$-th and $b$-th positions, at which $k_a$ and $k_b$ are switched; and
\begin{align}
  \tilde g_k
  \equiv \sum_{\substack{\set{i,j}\in\C_N\p{2}\\i\in k~\t{or}~j\in k}} g_{ij}
  = \sum_{i\in k} g_i - \sum_{\set{a,b}\in\C_M\p{2}} g_{k_ak_b}.
\end{align}
If the tensor $w$ is permutationally symmetric, meaning that $w_k$ is invariant under arbitrary permutations of $k$, then this symmetry is preserved by $g^{(M)}$.
In this case, we can replace sums over $k\in\D_N\p{M}$ in Eqs.~\eqref{eq:two_body_mat} and \eqref{eq:multi_body_mat} by sums over $k\in\C_N\p{M}$, and replace vectors $\ket{k_1,k_2,\cdots,k_M}\to\ket{\set{k_1,k_2,\cdots,k_M}}$, such that e.g.~$\ket{k_{a\leftrightarrow b}}=\ket{k}$.
These replacements reduce the size of $g^{(M)}$ from $\abs{\D_N\p{M}}\times\abs{\D_N\p{M}}$ to $\abs{\C_N\p{M}}\times\abs{\C_N\p{M}}$, where $\abs{\D_N\p{M}}=\prod_{j=0}^{M-1}\p{N-j}=M!\times{N\choose M}$ and $\abs{\C_N\p{M}}={N\choose M}$.
Additional symmetries of $g$ and $w$, such as translational invariance or lattice symmetries, can be used to further reduce the computational complexity of the eigenvalue problem in Eq.~\eqref{eq:multi_body_problem}.

\subsection{Recovering spin-wave theory}
\label{sec:spin_wave}

If the interaction Hamiltonian $\H_0$ is translationally invariant, then the single-body eigenvalue problem in Eq.~\eqref{eq:single_body_eig} is solvable analytically.
In this case, the couplings $g_{ij}$ depend only on the separation $\abs{i-j}$, so eigenvectors of $g$ are plane waves of the form
\begin{align}
  w_k \equiv \sum_{d\in\ZZ_L^D} e^{\i d\cdot k} \ket{d},
\end{align}
where on a $D$-dimensional periodic lattice of $N=L^D$ spins, lattice sites are indexed by vectors $d\in\ZZ_L^D$, and wavenumbers take on values $k\in\ZZ_L^D\times2\pi/L$.
The eigenvalues of $g$ can be determined by expanding
\begin{align}
  g \cdot w_k
  = \sum_{c,d\in\ZZ_L^D} g_{cd} e^{\i d\cdot k} \ket{c}
  = \sum_{c,d\in\ZZ_L^D} g_{c,c+d} e^{\i\p{c+d}\cdot k} \ket{c}
  = \sum_{d\in\ZZ_L^D} g_{0,d} \cos\p{d\cdot k} w_k,
\end{align}
where the imaginary contributions vanish in the sum over $d$ because $g_{0,d}=g_{0,-d}$.
The remainder of Eq.~\eqref{eq:single_body_eig} that we need to sort out is $\diag\vec g$, where all $g_i = \sum_{i,j}g_{ij} = \sum_d g_{0,d}$ are equal, which implies that $\diag\vec g = \sum_d g_{0,d}$ is a scalar.
We thus find that
\begin{align}
  g^{(1)} \cdot w_k = \Delta_k w_k,
  &&
  \Delta_k \equiv \sum_{d\in\ZZ_L^D} g_{0,d} \sp{\cos\p{d\cdot k}-1},
\end{align}
in agreement with standard spin-wave theory.
Excitations generated by the action of $\O\p{w_k,X}$ on PS states $\ket\psi\in\E_0$ are known as spin-waves.
If $g_{ij}=-U/N$ is constant, then the spin-wave excitation energies are $\Delta_k=U$ independent of the wavenumber $k$.

\section{Restricting spin operators to the permutationally symmetric manifold}
\label{sec:PS_ops}

Here we provide the restriction of a general $M$-body spin operator $\O$ to the permutationally symmetric (PS) manifold of $N$ spins (each with $n$ internal states).
Denoting the projector onto the PS manifold by $\P_0$, our task is essentially to find the coefficients of the expansion
\begin{align}
  \P_0 \O_M \P_0 = \sum_{a,b\in\A_n\p{N}} \bk{a|\O_M|b} \op{a}{b},
\end{align}
where $\A_n\p{N}$ is the set of all ways to assign $N$ (identical) spins to $n$ (distinct) states, such that for any $a\in\A_n\p{N}$ the state $\ket{a}=\ket{a_1,a_2,\cdots,a_n}$ is labeled by the occupation number $a_\mu$ of state $\mu$, with $\sum_\mu a_\mu=N$.
Written out explicitly,
\begin{align}
  \ket{a} = \f1{\sqrt{\C\p{a}}}
  \sum_{\substack{\t{distinct}\\\t{permutations}\\\hat\Pi\,\t{of}\,\tilde a}}
  \hat\Pi\ket{\tilde a},
  &&
  \ket{\tilde a} \equiv \bigotimes_\mu \ket{\mu}^{\otimes a_\mu},
  &&
  \C\p{a} \equiv \f{\p{\sum_\mu a_\mu}!}{\prod_\nu a_\nu!}.
\end{align}
Here $\C\p{a}$ is a multinomial coefficient that counts the number of distinct ways to permute the tensor factors of the ``standard-ordered'' state $\ket{\tilde a}$, enforcing $\bk{a|a}=1$.
Using these states, with some combinatorics we can expand
\begin{align}
  \bk{a|\O_M|b} =
  \sum_{\substack{\alpha,\beta\in\A_n\p{M}\\\alpha\le a,\beta\le b}}
  \delta_{a-\alpha,b-\beta}
  \sqrt{\f{\C\p{\alpha}\C\p{a-\alpha}\C\p{\beta}\C\p{b-\beta}}
    {\C\p{a}\C\p{b}}}
  \, \bk{\alpha|\O_M|\beta},
  \label{eq:multi_body_eval}
\end{align}
where the restriction $\alpha\le a$ and the difference $a-\alpha$ are evaluated element-wise, i.e.~$\alpha\le a\implies\alpha_\mu\le a_\mu$ and $\p{a-\alpha}_\mu=a_\mu-\alpha_\mu$ for all $\mu$; and $\delta_{cd}=1$ if $c=d$ and zero otherwise.
We sum over both $\alpha$ and $\beta$ above merely to keep the expression symmetric with respect to transposition $\p{a,\alpha}\leftrightarrow\p{b,\beta}$; in practice, one can simply sum over $\alpha\in\A_n\p{M}$ and set $\beta=b-a+\alpha$, throwing out terms with any $\beta_\mu<0$.
Note that, by slight abuse of notation, the operator $\O_M$ on the left of Eq.~\eqref{eq:multi_body_eval} acts on an arbitrary choice of $M$ spins (out of $N$), whereas the operator $\O_M$ on the right of Eq.~\eqref{eq:multi_body_eval} is simply an $M$-spin operator, with matrix elements $\bk{\alpha|\O_M|\beta}$ evaluated with respect to the PS $M$-spin states $\ket\alpha,\ket\beta\in\A_n\p{M}$.

\section{Relaxing assumptions of the three-laser drive}
\label{sec:full_drive}

In order to arrive at the drive Hamiltonian in Eq.~\eqref{eq:H_3LD_single} of the main text, we made two simplifying assumptions:
\begin{enumerate*}
\item that the excited-state hyperfine manifold had the same total spin $s$ as the ground-state manifold, and
\item that all drive amplitudes are real (which enforces a phase-locking condition between the driving lasers).
\end{enumerate*}
To derive an effective drive Hamiltonian for the general case in which the excited-state hyperfine manifold has total spin $s+r$ with $r\in\set{+1,0,-1}$, we decompose all lasers into their right- and left-circular polarization components and write the full drive Hamiltonian in the form
\begin{align}
  \H_{\t{drive}}^{\t{full}}
  = \sum_{j,\v v,\sigma} \Omega_{\v v\sigma}
  \p{e^{-\i\kappa\v v\cdot\v\ell j} \s_{\v v\sigma j}^{(r)}
    \otimes\op{\e}{\g}_j + \t{h.c.}}
  + \Delta \hat N_\e,
\end{align}
where $\Omega_{\v v\sigma}$ is the amplitude of $\sigma$-polarized light propagating along axis $\v v$, with $\sigma=+1$ and $-1$ respectively for right and left circular polarizations; and $\s_{\v v\sigma j}$ is a spin-raising/lowering operator for atom $j$ along axis $\v v$, defined by appropriately rotating the single-atom spin operators
\begin{align}
  \s_\pm^{(r)}
  \equiv -\sqrt{\f{n(n+1)(n-1)}{6}} \times \hat T_\pm^{(r)},
  &&
  \hat T_\pm^{(r)} \equiv \mp\sqrt{\f{2(s+r)+1}{2\ell+1}}
  \sum_\mu \bk{s\mu;1,\pm1|s+r,\mu\pm1} \op{\mu\pm1}{\mu}.
\end{align}
Here $\bk{j_1m_1;j_2m_2|j_3m_3}$ is a Clebsch-Gordan coefficient, and we have normalized $\hat T_\pm^{(r)}$ such that $\tr\sp{\hat T_\pm^{(r)}{}^\dag \hat T_\pm^{(r)}}=1$.
Still assuming real drive amplitudes, the corresponding effective drive Hamiltonian that replaces Eq.~\eqref{eq:H_3LD_single} in the far-detuned limit $\abs{\Delta}\gg\abs{\Omega_{\v v\sigma}}$ is then
\begin{align}
  \H_{\t{3LD}}^{\t{single}}
  = f_r^{(1)} \sp{\tilde\Omega_+ \tilde\Omega_- \s_\z
    + \tilde\Omega_0 \tilde\Omega_- \s_\x}
  + f_r^{(2)} \sp{\tilde\Omega_0 \tilde\Omega_+
    (\s_\z \s_\x  + \s_\x \s_\z)
    - \p{\tilde\Omega_0^2 \s_\z^2
      + \tilde\Omega_+^2 \s_\x^2
      + \tilde\Omega_-^2 \s_\y^2}}
  - f_r^{(3)} \sum_m \tilde\Omega_m^2,
\end{align}
where $f_r^{(k)}$ are scalars that depend on the spin dimension $n$:
\begin{align}
  f_0^{(1)} &= 1, &
  f_{+1}^{(1)} &= -s, &
  f_{-1}^{(1)} &= s+1, \\
  f_0^{(2)} &= 1, &
  f_{+1}^{(2)} &= -\f{s}{n+2}, &
  f_{-1}^{(2)} &= -\f{s+1}{n-2}, \\
  f_0^{(3)} &= 0, &
  f_{+1}^{(3)} &= \f{s(s+1)^2}{n+2}, &
  f_{-1}^{(3)} &= \f{s^2(s+1)}{n-2}.
\end{align}
If additionally the drive amplitudes are complex, $\Omega_m\to\Omega_me^{-\i\eta_m}$ (with real $\Omega_m,\eta_m$), then
\begin{multline}
  \H_{\t{3LD}}^{\t{single}}
  = f_r^{(1)} \tilde\Omega_+ \tilde\Omega_- \s_\z
  + \tilde\Omega_0 \sum_{\sigma\in\set{\pm1}}
  \f{\tilde\Omega_+ + \sigma\tilde\Omega_-}{2}
  \sp{f_r^{(1)} \sigma \s_{\tilde\eta_\sigma,\x}
    + f_r^{(2)} \p{\s_\z \s_{\tilde\eta_\sigma,\x}
      + \s_{\tilde\eta_\sigma,\x} \s_\z}} \\
  - f_r^{(2)} \sp{\tilde\Omega_0^2 \s_\z^2
    + \tilde\Omega_+ \s_{\tilde\eta_0,\x}^2
    + \tilde\Omega_- \s_{\tilde\eta_0,\y}^2}
  - f_r^{(3)} \sum_m \tilde\Omega_m^2,
\end{multline}
where $\s_{\eta\alpha} \equiv e^{-\i\eta \s_\z} \s_\alpha e^{\i\eta \s_\z}$ is a rotated spin-$\alpha$ operator (e.g.~$\s_{\pi/2,\x}=\s_\y$), and
\begin{align}
  \tilde\eta_\pm \equiv \pm \p{\eta_\pm - \eta_0},
  &&
  \tilde\eta_0 \equiv \f{\eta_+ - \eta_-}{2},
\end{align}
are the relative phases of the drive amplitudes.

\section{Mean-field theory}
\label{sec:MFT}

Here we describe the mean-field theory used to simulate the spin Hamiltonian
\begin{align}
  \H_\spin = -\f{u}{2N}\v\S\cdot\v\S + 2J\phi \sum_q \sin\p{q} \s_{\z,q}
\end{align}
in Eq.~\eqref{eq:H_spin} of the main text.
We begin by decomposing individual spin operators into Schwinger bosons as $\s_{\mu\nu q} = \b_{\mu q}^\dag \b_{\nu q}$, such that the spin Hamiltonian becomes
\begin{align}
  \H_\spin \to \H_{\t{boson}}
  = -\f{u}{2N} \sum_{p,q,\mu,\nu}
  \b_{\mu p}^\dag \b_{\nu p} \b_{\nu q}^\dag \b_{\mu q}
  + 2J\phi \sum_{q,\mu} \sin\p{q} \mu\, \b_{\mu q}^\dag \b_{\mu q}.
\end{align}
The Heisenberg equations of motion for the Schwinger boson operators are (see Appendix \ref{sec:bosons})
\begin{align}
  \i \partial_t \b_{\mu q}
  = -\f{u}{N} \sum_{\nu,p} \b_{\nu p}^\dag \b_{\mu p} \b_{\nu q}
  + 2J\phi \sin\p{q} \mu\, \b_{\mu q}.
\end{align}
Our mean-field theory then treats all boson operators in these equations of motion as complex numbers, $\b_{\mu q}\to\bk{\b_{\mu q}}_\MF$, with the initial value $\bk{\b_{\mu q}\p{t=0}}_\MF$ equal to the initial amplitude of spin $q$ in state $\mu$.
Specifically, for an $N$-fold product state of the form $\ket\psi=\bigotimes_q\sum_\mu\psi_{\mu q}\ket{\mu}$ we set $\bk{\b_{\mu q}\p{t=0}}_\MF = \psi_{\mu q}$.
For pure initial product states, this mean-field treatment of the boson operators $\b_{\mu q}$ is mathematically equivalent to a mean-field treatment of the spin operators $\s_{\mu\nu q}$, as in Eq.~\eqref{eq:H_MF}, but reduces the number of variables to keep track of by a factor of $\sim n$.

\section{Schwinger boson equations of motion for quadratic spin Hamiltonians}
\label{sec:bosons}

Here we decompose a quadratic spin Hamiltonian into Schwinger bosons, and derive the equations of motion for the resulting boson operators.
We begin with a general spin Hamiltonian of the form
\begin{align}
  \H = \sum_{\substack{\mu,\nu,\rho,\sigma\\j<k}}
  g^{\mu\nu j}_{\rho\sigma k} \s_{\mu\nu j} \s_{\rho\sigma k}
  + \sum_{\mu,\nu,j} \epsilon_{\mu\nu j} \s_{\mu\nu j},
  \label{eq:quadratic_spin}
\end{align}
where $\mu,\nu$ index orthogonal states of an $n$-level spin; $j,k$ index one of $N$ spins; $g^{\mu\nu j}_{\rho\sigma k}$ and $\epsilon_{\mu\nu j}$ are scalars; and $\s_{\mu\nu j}=\op{\mu}{\nu}_j$ is a transition operator for spin $j$.
Strictly speaking, Eq.~\eqref{eq:quadratic_spin} only defines the couplings $g^{\mu\nu j}_{\rho\sigma k}$ for $j<k$, so we enforce $g^{\mu\nu k}_{\rho\sigma j}=g^{\mu\nu j}_{\rho\sigma k}$ and $g^{\mu\nu j}_{\rho\sigma j}=0$ for completion.
Decomposing spin operators into Schwinger bosons as $\s_{\mu\nu j}=\b_{\mu j}^\dag \b_{\nu j}$, where $\b_{\nu j}$ a annihilates a boson of type $\nu$ on site $j$, we can write this Hamiltonian as
\begin{align}
  \H = \sum_{\substack{\mu,\nu,\rho,\sigma\\j<k}}
  g^{\mu\nu j}_{\rho\sigma k}
  \b_{\mu j}^\dag \b_{\nu j} \b_{\rho k}^\dag \b_{\sigma k}
  + \sum_{\mu,\nu,j} \epsilon_{\mu\nu j} \b_{\mu j}^\dag \b_{\nu j}.
\end{align}
The Heisenberg equations of motion for the boson operators are then
\begin{align}
  \i \partial_t \b_{\alpha\ell} = \sp{\b_{\alpha\ell}, \H}
  &= \sum_{\substack{\mu,\nu,\rho,\sigma\\j<k}}
  g^{\mu\nu j}_{\rho\sigma k}
  \sp{\b_{\alpha\ell}, \b_{\mu j}^\dag \b_{\nu j} \b_{\rho k}^\dag \b_{\sigma k}}
  + \sum_{\mu,\nu,j} \epsilon_{\mu\nu j}
  \sp{\b_{\alpha\ell}, \b_{\mu j}^\dag \b_{\nu j}} \\
  &= \sum_{\mu,\nu,\rho,\sigma,k} g^{\mu\nu\ell}_{\rho\sigma k}
  \sp{\b_{\alpha\ell}, \b_{\mu\ell}^\dag \b_{\nu\ell}}
  \b_{\rho k}^\dag \b_{\sigma k}
  + \sum_{\mu,\nu} \epsilon_{\mu\nu\ell}
  \sp{\b_{\alpha\ell}, \b_{\mu\ell}^\dag \b_{\nu\ell}} \\
  &= \sum_{\mu,\nu} \p{\sum_{\rho,\sigma,k}
    g^{\mu\nu\ell}_{\rho\sigma k} \b_{\rho k}^\dag \b_{\sigma k}
    + \epsilon_{\mu\nu\ell}}
  \sp{\b_{\alpha\ell}, \b_{\mu\ell}^\dag \b_{\nu\ell}}
\end{align}
where
\begin{align}
  \sp{\b_{\alpha\ell}, \b_{\mu\ell}^\dag \b_{\nu\ell}}
  = \delta_{\alpha\mu} \delta_{\alpha\nu} \b_{\alpha\ell}
  + \delta_{\alpha\mu} \p{1-\delta_{\alpha\nu}} \b_{\nu\ell}
  = \delta_{\alpha\mu} \b_{\nu\ell},
\end{align}
so
\begin{align}
  \i \partial_t \b_{\alpha\ell}
  = \sum_\nu \p{\sum_{\rho,\sigma,k}
    g^{\alpha\nu\ell}_{\rho\sigma k} \b_{\rho k}^\dag \b_{\sigma k}
    + \epsilon_{\alpha\nu\ell}} \b_{\nu\ell}.
\end{align}
In the case of uniform SU($n$)-symmetric interactions of the form $\frac{g}{2}\v\S\cdot\v\S$ and a diagonal external field, we have
\begin{align}
  g^{\alpha\nu\ell}_{\rho\sigma k}
  = g \times \delta_{\alpha\sigma} \delta_{\nu\rho},
  &&
  \epsilon_{\alpha\nu\ell}
  = \epsilon_{\alpha\ell} \times \delta_{\alpha\nu}
\end{align}
so
\begin{align}
  \i \partial_t \b_{\alpha\ell}
  = g \sum_{\nu,k} \b_{\nu k}^\dag \b_{\alpha k} \b_{\nu\ell}
  + \epsilon_{\alpha\ell} \b_{\alpha\ell}.
\end{align}

\section{Lax vector analysis}
\label{sec:lax}

We start with the spin Hamiltonian
\begin{align}
  \H_\spin
  = -\f{u}{2N} \sum_{\mu,\nu} \S_{\mu\nu} \S_{\nu\mu}
  + 2J\phi \sum_q \sin\p{q} \s_{\z,q},
\end{align}
where $\S_{\mu\nu} = \sum_q \s_{\mu\nu q}$.
The single-body operators that appear in this Hamiltonian have squared norms
\begin{align}
  \tr\p{\s_{\mu\nu q}^\dag \s_{\mu\nu q}} = 1
  &&
  \t{and}
  &&
  \tr\p{\s_{\z,q}^\dag \s_{\z,q}}
  = \sum_\mu \mu^2
  = \f1{12} (n+1) n (n-1)
  \equiv \xi^2.
\end{align}
The Lax formulation (following Refs.~\cite{yuzbashyan2005nonequilibrium, yuzbashyan2006dynamical, yuzbashyan2006relaxation, yuzbashyan2015quantum, smale2019observation}) requires all single-body operators involved to have the same normalization, so we substitute $\s_{\tilde\z,q}\equiv \s_{\z,q}/\xi$ to expand
\begin{align}
  \f{\H_\spin}{u}
  = -\f1{2N} \sum_{\mu,\nu} \S_{\mu\nu} \S_{\nu\mu}
  + \xi h \sum_q \sin\p{q} \s_{\tilde\z,q},
  &&
  \t{where}
  &&
  h \equiv \f{2J\phi}{u}.
\end{align}
The intensive, dimensionless, $(n^2-1)$-component Lax vector $\vec\ell\p{z}$ associated with $\H_\spin$, which is defined with an auxiliary complex parameter $z$, has components
\begin{align}
  \ell_\alpha\p{z}
  = \f1N \sum_q \f{\s_{\alpha,q}}{z-\sin q}
  + \delta_{\alpha,\tilde\z} \, \xi h,
\end{align}
where $\alpha$ indexes elements of a basis $\set{\s_\alpha}$ of self-adjoint generators of SU($n$), with normalization $\tr\p{\s_\alpha^2}=1$.
The squared magnitude $\vec\ell\p{z}^2=\sum_\alpha\ell_\alpha\p{z}^2$ is a constant of motion (for any $z$), and its residues provide $N$ mutually commuting quantities whose weighted sum recovers $\H_\spin$.
When $n=2$, conservation of these residues provides sufficient dynamical constraints to make the spin system fully integrable.
In this case, dynamical behavior is governed by the roots of $\vec\ell\p{z}^2$, and the presence (or absence) of complex roots marks distinct dynamical phases of $\H_\spin$.
However, the size of Hilbert space grows with $n$, while the number of conserved quantities provided by the Lax analysis (namely, $N$) does not.
When $n>2$, there is therefore no guarantee that the roots of $\vec\ell\p{z}^2$ will similarly govern dynamical behavior.
In fact, a straightforward generalization of the Lax analysis to $n>2$ makes predictions that are inconsistent with the mean-field results in Figures \ref{fig:mean_mag_int_X}--\ref{fig:mean_mag_int_XXI} of the main text.
We substantiate this claim with a direct calculation of the roots of $\vec\ell\p{z}^2$ below.

Within the permutationally symmetric manifold, we can replace $\s_{\alpha,q}\to\bar s_\alpha\equiv\frac1N\sum_q \s_{\alpha,q}$ at the cost of $O(1/N)$ errors that vanish as $N\to\infty$, so taking this limit we find
\begin{align}
  \ell_\alpha\p{z}
  = \I\p{z} \bar s_\alpha
  + \delta_{\alpha,\tilde\z} \, \xi h,
\end{align}
where
\begin{align}
  \I\p{z} \equiv \lim_{N\to\infty} \f1N \sum_q \f1{z-\sin\p{q}}
  = \f1{2\pi} \int_0^{2\pi} \f{\dd q}{z-\sin\p{q}}
  = \f1{\sqrt{z^2-1}}
  &&
  \t{for}
  &&
  z \notin \sp{-1,1}.
\end{align}
The squared magnitude of the Lax vector is therefore
\begin{align}
  \vec\ell\p{z}^2
  = \sum_\alpha \ell_\alpha\p{z}^2
  = \I\p{z}^2 \sum_{\alpha\ne\tilde\z} \bar s_\alpha^2
  + \sp{\I\p{z} \bar s_{\tilde\z} + \xi h}^2,
\end{align}
where we can define the scalar $Q^2\equiv\sum_\alpha \bar s_\alpha^2$ to simplify
\begin{align}
  \vec\ell\p{z}^2
  = \I\p{z}^2 \p{Q^2 - \bar s_{\tilde\z}^2}
  + \sp{\I\p{z} \bar s_{\tilde\z} + \xi h}^2
  = \I\p{z}^2 Q^2 + \xi^2 h^2
  + 2 \I\p{z} \xi h \bar s_{\tilde\z}.
\end{align}
For initial states with $\bk{\bar s_\z}=0$, we thus find that
\begin{align}
  \vec\ell\p{z}^2 = \f{Q^2}{z^2-1} + \xi^2 h^2,
\end{align}
which is zero when\footnote{Strictly speaking, the zeros in Eq.~\eqref{eq:lax_zeros} occur at values of $z$ at which $\I\p{z}$ is undefined.
We avoid this issue by analytically continuing $\I\p{z}^2$ to the interval $z\in(-1,1)$.}
\begin{align}
  z = \pm \sqrt{1 - \p{\f{Q}{\xi h}}^2}.
  \label{eq:lax_zeros}
\end{align}
These roots change character when $z=0$, suggesting that the critical field $h_\crit$ separating dynamical phases satisfies
\begin{align}
  h_\crit^2 \stackrel{?}{=} \f{Q^2}{\xi^2},
\end{align}
where we use the relation $\stackrel{?}{=}$ to indicate that this ``prediction'' of the Lax analysis is not necessarily valid for all $n$.
For a permutationally symmetric state, up to vanishing $O(1/N)$ corrections we can expand
\begin{align}
  Q^2 = \sum_\alpha \bar s_\alpha^2
  = \sum_{\mu,\nu} \bar s_{\mu\nu} \bar s_{\nu\mu} - \f1n
  = 1 - \f1n
  = \f{n-1}{n},
\end{align}
which implies that
\begin{align}
  h_\crit^2 \stackrel{?}{=} \f{n-1}{n} \times \f{12}{n(n+1)(n-1)}
  = \f{12}{n^2\p{n+1}}.
\end{align}
This Lax analysis correctly predicts that $h_\crit=1$ when $n=2$, but otherwise predicts $h_\crit\sim n^{-3/2}$, which is inconsistent with the finding that $h_\crit\sim n^{-1/3}$ in the mean-field results of the main text (see Figure \ref{fig:crit_fields_X}).
We emphasize that this inconsistency is not a failure of the Lax formalism, but rather an indication that new theoretical tools are necessary to understand multilevel spin models.

\end{document}